\documentclass[reprint, amsmath,amssymb,aps,pra]{revtex4-1}

\usepackage{amsmath}
\usepackage{amssymb}
\usepackage{graphicx}
\usepackage{subfig}
\usepackage{color}
\usepackage{dcolumn}% Align table columns on decimal point
\usepackage{bm}% bold math
\begin{document}

%\preprint{APS/123-QED}

\title{A simple, analytic solution of the Rabi model using multiple-scales}

\author{L. O. Casta\^{n}os}

\email{LOCCJ@yahoo.com, luis.castanos@itesm.mx}
\affiliation{Instituto Tecnol\'{o}gico y de Estudios Superiores de Monterrey, Campus Santa Fe, Avenida de los Poetas 100, Santa Fe, La Loma, 01389 Ciudad de M\'{e}xico, M\'{e}xico}%

\date{\today}

\begin{abstract}
We considered the semiclassical Rabi model, that is, a two-level system interacting with a single-mode, classical field. In resonance and for an arbitrary initial state of the system, we obtained a simple, approximate, and analytic solution that takes into account the counterrotating terms and that is accurate as long as the Rabi frequency is smaller than or equal to $1/2$ ($1/4$) of the angular frequency of the field and $1$ ($10$) Rabi oscillations are considered. In addition, the approximate solution has the same level of complexity as that obtained with the rotating-wave-approximation (RWA) and allows one to describe the evolution of the Bloch vector of the system in terms of a \textit{slow-precessional} and a \textit{fast-nutational} motion similar to that of a symmetric top with one point fixed. Finally, the approximate solution also leads to a simple criterion that allows one to determine the accuracy of the solution in the RWA.
\end{abstract}

\pacs{03.65.-w, 42.50.-p, 31.15.xg, 03.65.Aa}

\keywords{semiclassical Rabi model, semiclassical radiation theory, two-level system, classical field, counterrotating terms}
\maketitle

\section{Introduction}
  
The semiclassical and quantum Rabi models are composed of a two-level system interacting with a  classical or quantum single-mode field, respectively. They constitute two fundamental models used to describe matter-light interaction \cite{80} where the two-level system models either a real or an artificial atom and the field describes a single-mode electromagnetic field. In many cases, such as those considered in the interaction of a real atom with a nearly resonant, single-mode electromagnetic field,  both the coupling between the two-level system and the field and the detuning between the two-levels' transition frequency and the field's frequency are much smaller than the field's frequency. In these cases the two models can be simplified using the \textit{rotating-wave-approximation} (RWA) where \textit{counterrotating terms} are neglected. The results are simple models that can be solved exactly and that have been used extensively to describe many physical phenomena such as those found in cavity quantum electrodynamics (QED) experiments \cite{libro,Maser1,Maser2,Shakov}. For example, the semiclassical Rabi model with the rotating wave approximation has been used to describe how a two-level atom can be prepared or analyzed in a state that corresponds to an arbitrary point on the surface of the Bloch sphere, while the quantum Rabi model with the RWA (the Jaynes-Cummings model) has been used to describe the interaction between a two-level atom and a single-mode quantum electromagnetic field in maser experiments. Nevertheless, there are many systems where the RWA is not valid and one must consider the full Rabi models. Therefore, investigating the properties of these models and their extensions is important for the adequate understanding of the physics of these more complicated systems. This is specially relevant with the advent of experimental systems that can access a coupling strength and/or a detuning such that the RWA is no longer applicable \cite{Devoret,Niem,Forn,Ion}. In particular, this is the case of the area of circuit QED \cite{Devoret,Niem,Forn}. In fact, experiments in this area described in \cite{Niem,Forn} have already explicitly shown the breakdown of the RWA. 

Actually, the semiclassical and quantum Rabi models are quite difficult to solve exactly. In fact, the exact solution of the quantum Rabi model was found until quite recently \cite{Braak,Nuevo1,Nuevo2,Nuevo3}. In addition, several approximate treatments have also been developed. Among these, there is are an adiabatic treatment \cite{IrishPRB}, a generalized RWA \cite{Irish},  the use of van Vleck perturbation theory \cite{Hausinger}, and a generalized variational method \cite{Zhang}. These approximate treatments can be used to describe the system in regimes where the RWA is no longer valid and have the advantage of usually providing simpler expressions for the energies and the corresponding eigenstates than the exact solution of the quantum Rabi model. Moreover, there have also been studies of the dynamics of a two-level system interacting with a quantum field in a regime beyond the RWA  \cite{Casanova}, as well as the possibility of preparing nonclassical states in this system \cite{Ashab}. In addition, the dynamics and entanglement beyond the RWA of a system composed of two and three two-level systems interacting with a quantum field have also been investigated \cite{Eberly, Zhang, Zhang2}.

In this article we considered the semiclassical Rabi model, that is, we considered a two-level system interacting with a single-mode, classical field. The objective was to obtain an analytic, approximate solution that describes the evolution of the density operator of the system, that includes the effect of the counterrotating terms, that has the same level of simplicity as the one obtained with the RWA, and that is accurate even when the coupling with the field is comparable with the field's frequency. The motivation for this is that such analytic, approximate solution could be used to describe the system in a parameter regime where the RWA does not hold and that it could be used both to achieve a better understanding of the effects of the counterrotating terms and to estimate the error made when using the solution with the RWA. In this article we obtained an analytic, approximate solution that satisfies in great measure all of the aforementioned properties when the field is resonant with the two-level's transition. 

The article is organized as follows. In Section II we introduce the model describing the system under consideration and in Section III we present the density operator describing the evolution of the system with the RWA. In Section IV we establish the approximate density operator describing the evolution of the system taking into account the counterrotating terms neglected in the RWA. In addition, we also determine the accuracy of the approximation and establish a criterion that can be used to determine the accuracy of the RWA. Finally, a summary and the conclusions are given in Sec. V.

\section{The model}

We consider a two-level system (a two-level real or artificial atom or a qubit) interacting with a single-mode, classical field of angular frequency $\omega_{1} >0$. We assume that the field is resonant with the angular transition frequency of the two-level system. Let $\mathcal{H}_{A}$ denote the state space of the system and let $\rho (t)$ be the density operator of the system at time $t$. An orthonormal basis for $\mathcal{H}_{A}$ is $\gamma = \{  \ \vert 1 \rangle , \ \vert 2 \rangle  \ \}$, where $\vert 1 \rangle$ is the ground state of the qubit and $\vert 2\rangle$ is the excited state. Hence, one has the following closure and orthonormalization relations:
\begin{eqnarray}
\label{1Base}
\vert 1 \rangle\langle 1 \vert + \vert 2 \rangle\langle 2 \vert = \mathbf{I} \ , \quad \langle \lambda \vert \lambda ' \rangle = \delta_{\lambda \lambda '} \quad (\lambda, \lambda' = 1,2).
\end{eqnarray}
Here $\mathbf{I}$ is the identity operator in $\mathcal{H}_{A}$, and $ \delta_{\lambda \lambda '}$ is the Kronecker delta.

We assume that the Hamiltonian of the system is 
\begin{eqnarray}
\label{2Hamiltoniano}
H(t) &=& H_{A} + H_{I}^{0}(t)   \ ,   
\end{eqnarray}
where $H_{A}$ is the Hamiltonian of the qubit
\begin{eqnarray}
\label{2HamiltonianoB}
H_{A} = \frac{\hbar \omega_{1}}{2}\left( \ \vert 2 \rangle\langle 2 \vert - \vert 1 \rangle\langle 1 \vert   \ \right) \  ,
\end{eqnarray}
and $H_{I}^{0}(t)$ is the interaction Hamiltonian between the qubit and the field
\begin{eqnarray}
\label{4H}
H_{I}^{0}(t) &=& -\frac{\hbar \Omega_{0}}{2}\left( e^{i\omega_{1}t} + e^{-i\omega_{1}t}   \right) \left( b + b^{\dagger} \right) \ .
\end{eqnarray}
Here $\Omega_{0} >0$ is the Rabi frequency, a quantity with units $1/s$ that determines the magnitude of the qubit-field coupling, and
\begin{eqnarray}
\label{4d}
b = \vert 1 \rangle\langle 2 \vert  \ .
\end{eqnarray}
Notice that the Hamiltonian in (\ref{2Hamiltoniano})-(\ref{4d}) corresponds to the semiclassical Rabi model with the condition of resonance. It can describe, for example, a two-level atom fixed at a position and interacting in the electric dipole approximation with an electric field that is linearly polarized. In this case, the two levels of the atom correspond to states with definite parity and to a $\Delta m = 0$ transition, see \cite{Mandel} for the origin of (\ref{2Hamiltoniano})-(\ref{4d}) in this case.

We now pass to the interaction picture (IP) defined by the unitary operator
\begin{eqnarray}
\label{11}
U_{I}(t) = \mbox{Exp}\left[ \ {-\frac{i}{\hbar}}H_{A}t  \ \right] \ .
\end{eqnarray}
For clarity, in the following a subindex $I$ in an operator $A_{I}(t)$ will indicate that the operator is in the IP, that is,
\begin{eqnarray}
\label{12}
A_{I}(t) = U_{I}^{\dagger}(t) A(t) U_{I}(t) \ ,
\end{eqnarray} 
where $A(t)$ is a linear operator in $\mathcal{H}_{A}$ in the Sch\"{o}dinger picture.

It follows that the equation governing the evolution of the density operator $\rho (t)$ of the system in the IP  (von Neumann's equation in the IP) is given by
\begin{eqnarray}
\label{13}
i\hbar \frac{d}{dt} \rho_{I}(t) = \left[  H_{II}^{0}(t), \rho_{I}(t)  \right] \ , \
\end{eqnarray}
where $[ \cdot , \cdot ]$ is the commutator and $H_{II}^{0}(t)$ is the interaction Hamiltonian in the IP
\begin{eqnarray}
\label{15}
H_{II}^{0}(t) &=& -\frac{\hbar\Omega_{0}}{2}\left( b + b^{\dagger}   \right) -\frac{\hbar\Omega_{0}}{2}\left( e^{-i2\omega_{1}t}b + e^{i2\omega_{1}t}b^{\dagger}   \right) .
\end{eqnarray}
In (\ref{15}) one immediately recognizes $H_{II}^{0}(t)$ as being the sum of a time-independent part (the resonant or rotating terms) and a time-dependent part (the non resonant or counterrotating  terms).

In the next sections we solve von Neumann's equation (\ref{13}) to good approximation. To present the results it is convenient to introduce the matrix representation $[ \rho_{I}(t)]_{\gamma}$ of the IP density operator $\rho_{I}(t)$ in the basis $\gamma = \{ \ \vert 1 \rangle, \ \vert 2 \rangle \ \}$. It is given by
\begin{eqnarray}
\label{Matriz}
\left[ \rho_{I}(t) \right]_{\gamma} &=& \left(
\begin{array}{cc}
\rho_{11} (t) & \rho_{12} (t) \cr
\rho_{21} (t) & \rho_{22} (t) 
\end{array}
\right) \ , \cr 
&& \cr
&=& 
\left(
\begin{array}{cc}
\frac{1}{2}[ 1 - \alpha_{30} (t) ]  & \rho_{12} (t) \cr
\rho_{12} (t)^{*} & \frac{1}{2}[ 1 + \alpha_{30} (t) ] 
\end{array}
\right) \ ,
\end{eqnarray}
with 
\begin{eqnarray}
\label{Matriz2}
\rho_{\lambda \lambda'}(t) &=& \langle \lambda \vert \rho_{I}(t) \vert \lambda' \rangle \ , \quad (\lambda , \lambda ' = 1,2) .
\end{eqnarray}
Notice that in (\ref{Matriz}) we used that 
\begin{eqnarray}
\label{Matriz3}
\rho_{11}(t) &=& 1-\rho_{22}(t) \ , \quad \rho_{21}(t) = \rho_{12}(t)^{*} \  , \cr
\alpha_{30}(t) &=& \rho_{22}(t) -\rho_{11}(t) \ .
\end{eqnarray}
The first two equalities hold because Tr$[\rho_{I}(t)] = 1$ and $\rho_{I}(t)$ is Hermitian, while the third defines $\alpha_{30}(t)$. Here and in the following $z^{*}$ denotes the complex conjugate of the quantity $z$.

In order to determine the accuracy of the approximate analytic results, we also solve (\ref{13}) numerically. To facilitate the comparison between the analytical and numerical results and to give a geometric interpretation of them, it is convenient to introduce the \textit{Bloch vector} \cite{Mandel}. It is defined by
\begin{eqnarray}
\label{Bloch}
\mathbf{r}_{I}(t) &=& \Big( \ \alpha_{10}(t), \ \alpha_{20}(t), \ \alpha_{30}(t) \ \Big)^{T} \ ,
\end{eqnarray} 
where T denotes the transpose, $\alpha_{30}(t)$ is defined in (\ref{Matriz3}), and
\begin{eqnarray}
\label{ComponentesBloch}
\alpha_{10}(t) &=& 2 \mbox{Re} \left[ \  \rho_{12}(t) \ \right] \ , \ \  \alpha_{20}(t) = 2 \mbox{Im} \left[ \  \rho_{12}(t) \ \right] .
\end{eqnarray}
Here Re($\cdot$) and Im($\cdot$) denote the real and imaginary parts of a complex number, respectively.  Finally, we recall some properties of the Bloch vector \cite{Mandel} that are used in the following sections:
\begin{enumerate}
\item $\rho_{I} (t)$ is a pure state if and only if $\vert \mathbf{r}_{I}(0) \vert = 1$.
\item $\rho_{I} (t)$ is a mixed state if and only if $\vert \mathbf{r}_{I}(0) \vert < 1$.
\end{enumerate}
Here and in the following $\vert \cdot \vert$ denotes the Euclidean or $2$ norm.

%%%%%%%%%%%%%%%%%%%%%%%%%%%%%%%%%%
%%%%%%%%%%%%%%%%%%%%%%%%%%%%%%%%%%
%%%%%%%%%%%%%%%%%%%%%%%%%%%%%%%%%%
%%%%%%%%%%%%%%%%%%%%%%%%%%%%%%%%%%

\section{The solution in the rotating-wave-approximation (RWA)}

A rigorous mathematical deduction of the rotating-wave-approximation (RWA) is obtained by using the \textit{Averaging Theorem} \cite{GH}. Here we prefer to deduce it using physical arguments.

In the RWA one first observes that the interaction Hamiltonian $H_{II}^{0}(t)$ in (\ref{15}) includes time-dependent terms that vary as $e^{\pm i2\omega_{1}t}$ and, thus, that evolve appreciably in a time-scale $1/(2\omega_{1})$. Then, one assumes that the qubit-field coupling $\Omega_{0}$ is small and that $\rho_{I}(t)$ evolves appreciably in a time-scale much larger than $1/(2\omega_{1})$, so that the terms in (\ref{15}) that are multiplied by $e^{\pm i2\omega_{1}t}$ average to zero. Hence, one can neglect the terms in $H_{II}^{0}(t)$ that are multiplied by $e^{\pm i2\omega_{1}t}$ and one is led to the approximation 
\begin{eqnarray}
\label{15b}
H_{II}^{0}(t) &\simeq& -\frac{\hbar\Omega_{0}}{2}\left( b + b^{\dagger}   \right) \ .
\end{eqnarray}
Von Neumann's equation in  (\ref{13}) with the approximate interaction Hamiltonian in (\ref{15b}) can be solved exactly, see the appendix for the details. The exact solution is 
\begin{eqnarray}
\label{RWAsolucionE}
\rho_{12}^{\mbox{\tiny RWA}}(t) &=&  \rho_{12}(0) \mbox{cos}^{2}\left( \frac{\Omega_{0}t}{2} \right)  + \rho_{12}(0)^{*} \mbox{sin}^{2}\left( \frac{\Omega_{0}t}{2} \right)   \cr
&& + \alpha_{30}(0)   \frac{i}{2} \mbox{sin}\left( \Omega_{0}t \right) \  ,  \cr
&& \cr
\alpha_{30}^{\mbox{\tiny RWA}}(t) &=& i \rho_{12}(0) \mbox{sin}\left( \Omega_{0}t \right)  - i \rho_{12}(0)^{*} \mbox{sin}\left( \Omega_{0}t \right)  \cr
&& + \alpha_{30}(0) \mbox{cos}\left(\Omega_{0}t\right) .
\end{eqnarray}
Notice that we have included the superscript \textit{RWA} to indicate that it is the solution in the RWA.

To obtain the density matrix of the system in the basis $\gamma = \{ \vert 1 \rangle , \ \vert 2 \rangle \}$, one simply substitutes (\ref{RWAsolucionE}) in the righthand side of (\ref{Matriz}).

From (\ref{RWAsolucionE}) it follows that the matrix elements of $\rho_{I}(t)$ are periodic functions with period $2\pi/\Omega_{0}$ (the period of one \textit{Rabi oscillation}).  Then, $\rho_{I}(t)$ evolves appreciably in a time-scale of  $1/ \Omega_{0}$. Since to perform the RWA it was assumed that $\Omega_{0}$ is small and that $\rho_{I}(t)$ evolves on a time-scale much larger than $1/(2\omega_{1})$, it follows that the RWA holds if $\Omega_{0} \ll 2\omega_{1}$. This is the well-known result for the validity of the RWA in resonance \cite{CohenAPI}.

To end this section we use (\ref{RWAsolucionE}) to write the Bloch vector $\mathbf{r}_{I}^{\mbox{\tiny RWA}}(t)$ in the RWA:
\begin{eqnarray}
\label{TE36Bloch}
\mathbf{r}_{I}^{\mbox{\tiny RWA}}(t) &=& \mathcal{R}(\Omega_{0}t)\mathbf{r}_{I}(0) \ ,
\end{eqnarray}
with
\begin{eqnarray}
\label{TE36matriz}
\mathcal{R}(\tau) &=& \left(
\begin{array}{ccc}
1 & 0 & 0 \cr
0 & \mbox{cos}(\tau) & \mbox{sin}(\tau) \cr
0 & -\mbox{sin}(\tau) & \mbox{cos}(\tau) 
\end{array}
\right) .
\end{eqnarray}
Observe that $\mathcal{R}(\Omega_{0}t)$ is an orthogonal matrix that performs a rotation of an angle $-\Omega_{0}t$ around de $x$-axis. Therefore, in the RWA, the Bloch vector at time $t$ is obtained by rotating around the $x$-axis the Bloch vector at time $0$.  The rotation is of an angle $\Omega_{0}t$ in the clockwise sense  when  $t\geq 0$.

%%%%%%%%%%%%%%%%%%%%%%%%%%%%%%%%%%
%%%%%%%%%%%%%%%%%%%%%%%%%%%%%%%%%%
%%%%%%%%%%%%%%%%%%%%%%%%%%%%%%%%%%
%%%%%%%%%%%%%%%%%%%%%%%%%%%%%%%%%%

\section{The multiple-scales solution}

The results of the last section indicate that the RWA holds when there are two clearly separated time-scales in the system under consideration: a fast time-scale $1/(2\omega_{1})$ in which the counterrotating (or time-dependent qubit-field interaction) terms evolve appreciably and a slow time-scale $1/\Omega_{0}$ in which the IP density operator $\rho_{I}(t)$ evolves appreciably. Instead of neglecting the counterrotating terms, one can use the method of multiple-scales \cite{Holmes} to solve von Neumann's equation in the IP (\ref{13}) with the complete interaction Hamiltonian in (\ref{15}). This allows one to obtain a simple, approximate analytic solution that is accurate for long times. In addition, the multiple scales solution provides physical insight into the effect of the counterrotating terms and it allows one to determine both corrections to the RWA and a quantitative criterion that indicates when the RWA leads to accurate results. This is done below. 
%%%%%%%%%%%%%%%%%%%%%%%%%%%%%%%%%%
%%%%%%%%%%%%%%%%%%%%%%%%%%%%%%%%%%
%%%%%%%%%%%%%%%%%%%%%%%%%%%%%%%%%%
%%%%%%%%%%%%%%%%%%%%%%%%%%%%%%%%%%

\subsection{Geometric description}

In this section we first give a geometric interpretation of the evolution of the system. Afterwards, we introduce the multiple-scales method and use the geometric interpretation to explain what the multiple scales method is doing. In order to this, it is convenient to use the Bloch vector of the system defined in (\ref{Matriz3})-(\ref{ComponentesBloch}).

If one expresses von Neumann's equation in (\ref{13}) and (\ref{15}) in terms of the matrix elements $\rho_{\lambda \lambda'}(t)$ of the IP density operator and then one writes the resulting equations in terms of the components of the Bloch vector $\mathbf{r}_{I}(t)$, one obtains the following equation:
\begin{eqnarray}
\label{TE34}
\frac{d}{dt} \mathbf{r}_{I}(t) &=& -\Omega_{0} \left[ \mathbf{\hat{x}} +  \mathbf{\hat{w}}(2\omega_{1}t)  \right] \times \mathbf{r}_{I}(t)  ,
\end{eqnarray}
where $\times$ indicates the cross-product and
\begin{eqnarray}
\label{TE34w}
\mathbf{\hat{w}}(2\omega_{1}t) &=& \mathbf{\hat{x}}\mbox{cos}(2\omega_{1}t) - \mathbf{\hat{y}}\mbox{sin}(2\omega_{1}t) \  .
\end{eqnarray}
Here and in the following $\mathbf{\hat{x}}$, $\mathbf{\hat{y}}$, and $\mathbf{\hat{z}}$  denote unit vectors in the positive directions of the $x$-, $y$-, and $z$-axes, respectively. We note that (\ref{TE34}) is equivalent to von Neumann's equation in (\ref{13}) and (\ref{15}).

Observe that (\ref{TE34}) is the equation that governs the evolution of a point that is rotating \cite{Marion} around the unit vector 
\begin{eqnarray}
\label{TE34vR}
\mathbf{\hat{Q}}(t) &=& \frac{\mathbf{\hat{x}} + \mathbf{\hat{w}}(2w_{1}t)}{\vert \mathbf{\hat{x}} + \mathbf{\hat{w}}(2w_{1}t) \vert} \ ,
\end{eqnarray}
with angular velocity
\begin{eqnarray}
\label{TE34fR}
q(t) &=& -\Omega_{0}\vert \mathbf{\hat{x}} + \mathbf{\hat{w}}(2w_{1}t) \vert . 
\end{eqnarray}
Recall that a vector $\mathbf{r}_{1}$ rotates around another vector $\mathbf{r}_{2}$ with an angular velocity $\omega$ means that $\mathbf{r}_{1}$ rotates around $\mathbf{r}_{2}$ in the clockwise (counterclockwise) sense when viewed from the tip of $\mathbf{r}_{2}$ with an angular speed $\vert \omega \vert$ if $\omega <0$ ($\omega >0$). Since $q(t)<0$, it follows that the Bloch vector $\mathbf{r}_{I}(t)$ rotates around $\mathbf{\hat{Q}}(t)$ in the clockwise sense (as viewed from the tip of $\mathbf{\hat{Q}}(t)$) with an angular speed $\vert q(t)\vert$. 

We now give an interpretation of the evolution of the Bloch vector $\mathbf{r}_{I}(t)$. First, observe from (\ref{TE34w}) that the vector $\mathbf{\hat{w}}(2\omega_{1}t)$ is rotating in the $xy$-plane and around the $z$-axis in the clockwise sense (when viewed from the tip of $\mathbf{\hat{z}}$) with an angular speed $2\omega_{1}$ and that it takes a time $T = 2\pi/(2\omega_{1})$ for $\mathbf{\hat{w}}(2\omega_{1}t)$ to make one complete turn around the $z$-axis. Second, notice from (\ref{TE34}) that, as $\mathbf{\hat{w}}(2\omega_{1}t)$ rotates around the $z$-axis,  $\mathbf{r}_{I}(t)$ rotates around $\mathbf{\hat{w}}(2\omega_{1}t)$ with an angular velocity $-\Omega_{0}$ and $\mathbf{r}_{I}(t)$  also rotates around $\mathbf{\hat{x}}$ with an angular velocity $-\Omega_{0}$.

Assume that $\Omega_{0} \ll 2\omega_{1}$. Then,  $\mathbf{\hat{w}}(2\omega_{1}t) $ is rotating around the $z$-axis much faster than $\mathbf{r}_{I}(t)$ is rotating around $\mathbf{\hat{w}}(2\omega_{1}t)$ and $\mathbf{\hat{x}}$. As a consequence, $\mathbf{r}_{I}(t)$ changes only slightly during one complete turn of $\mathbf{\hat{w}}(2\omega_{1}t) $ around the $z$-axis. Also, during one complete turn of $\mathbf{\hat{w}}(2\omega_{1}t) $ around the $z$-axis, the rotation of the Bloch vector $\mathbf{r}_{I}(t)$ around $\mathbf{\hat{w}}(2\omega_{1}t)$ is going to \textit{approximately average to zero}. The reason for this is the following. Suppose that the complete turn goes from $t = t_{1}$ to $t = t_{1} + T$. Since $\mathbf{r}_{I}(t)$ changes only slightly during this time interval, the rotation of $\mathbf{r}_{I}(t)$ induced by $\mathbf{\hat{w}}(2\omega_{1}t)$  at any time $t = t_{2} \in [t_{1}, t_{1} +(T/2) )$  is going to be almost completely cancelled by the rotation induced by $\mathbf{\hat{w}}(2\omega_{1}t)$ at the later time $t = t_{2} + T/2 \in[t_{1}+(T/2), t_{1}+T)$ because $\mathbf{\hat{w}}(2\omega_{1}t)$ at time $t = t_{2} + T/2$ points in the direction opposite to that of $\mathbf{\hat{w}}(2\omega_{1}t)$ at time $t = t_{2}$. Moreover, the aforementioned cancellation is going to be more exact (that is, the difference between the two terms is going to tend to zero) for smaller $\Omega_{0}$ because $\mathbf{\hat{w}}(2\omega_{1}t) $ is going to rotate much faster  around the $z$-axis than $\mathbf{r}_{I}(t)$ is going to rotate around $\mathbf{\hat{w}}(2\omega_{1}t)$ and $\mathbf{\hat{x}}$.

From the discussion of the preceding paragraph one has that $\mathbf{r}_{I}(t)$ does not change appreciably during one complete turn of $\mathbf{\hat{w}}(2\omega_{1}t)$ around the $z$-axis and, as a consequence,  the rotation of $\mathbf{r}_{I}(t)$ around $\mathbf{\hat{w}}(2\omega_{1}t)$ and $\mathbf{\hat{x}}$ approximately reduces to a rotation of $\mathbf{r}_{I}(t)$ around $\mathbf{\hat{x}}$ with an angular velocity $-\Omega_{0}$. Hence,  one can neglect in (\ref{TE34}) the term multiplied by $ \mathbf{\hat{w}}(2\omega_{1}t)$ to obtain the approximate equation
\begin{eqnarray}
\label{TE34rwa}
\frac{d}{dt} \mathbf{r}_{I}(t) &=& -\Omega_{0} \mathbf{\hat{x}} \times \mathbf{r}_{I}(t)  .
\end{eqnarray}
We note that neglecting the term multiplied by $ \mathbf{\hat{w}}(2\omega_{1}t)$ corresponds to performing the rotating-wave-approximation (RWA). In other words, we have just given a geometric description of the RWA and (\ref{TE34rwa}) is equivalent to von Neumann's equation (\ref{13}) with the RWA interaction Hamiltonian in (\ref{15b}). Note that the term $-\Omega_{0}\mathbf{\hat{x}}$ in (\ref{TE34}) corresponds to the rotating (or resonant) terms in $H_{II}^{0}(t)$, while the term  $-\Omega_{0}\mathbf{\hat{w}}(2\omega_{1}t)$ in (\ref{TE34}) corresponds to the counterrotating (or nonresonant) terms in $H_{II}^{0}(t)$.

Observe that  (\ref{TE34rwa}) indicates that $\mathbf{r}_{I}(t)$ only rotates around $\mathbf{\hat{x}}$ in the clockwise sense with an angular speed $\Omega_{0} \ll 2\omega_{1}$. In other words, $\mathbf{r}_{I}(t)$ exhibits a \textit{slow precessional-motion} around the $x$-axis. Notice that this coincides with the interpretation of the evolution of the Bloch vector in the RWA given after equation (\ref{TE36matriz}).

\begin{figure}[htbp]
     \centering
     \subfloat[]{\label{Figure0a}\includegraphics[scale=0.8]{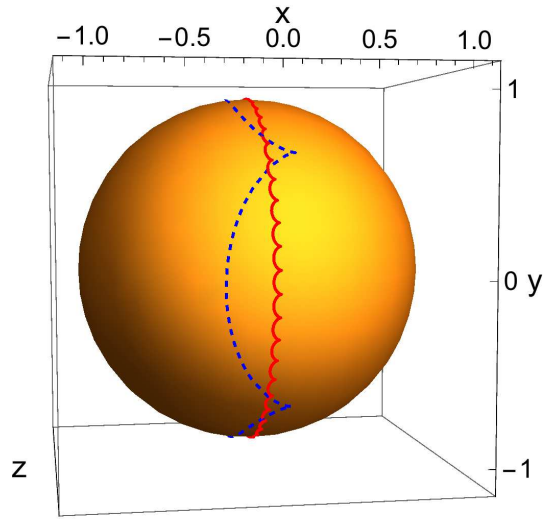}}\\
     \subfloat[]{\label{Figure0b}\includegraphics[scale=0.8]{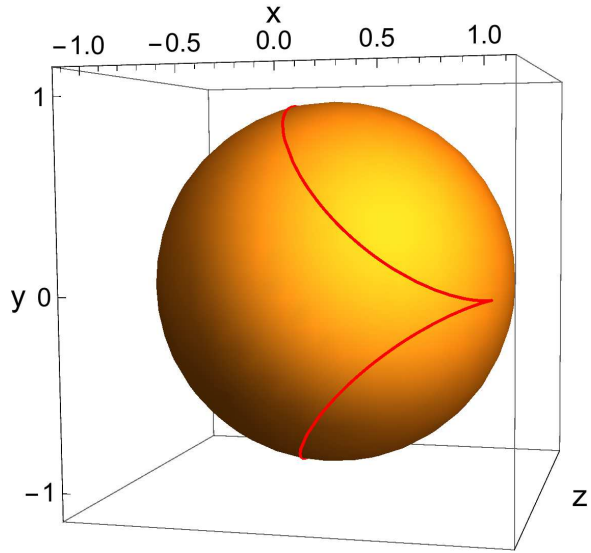}}
     \caption{(Color online) The figures show part of the trajectory followed by the Bloch vector in the Bloch sphere (the sphere of radius $1$ centered at the origin) for several values of $\epsilon = \Omega_{0}/(2\omega_{1})$. Figure \ref{Figure0a} illustrates the cases $\epsilon = 1/50$ (red-solid line) and $\epsilon = 1/4$ (blue-dashed line), while Figure \ref{Figure0b} shows the case $\epsilon = 1$ (red-solid line). In all figures the Bloch vector was calculated numerically by solving (\ref{TE34}) as a function of the non dimensional time $\tau = 2\omega_{1}t$ (explicitly, equations (\ref{eqBloch}) in the appendix) from $\tau = 0$ to $\tau = (2\pi/\epsilon)$ (that is, one Rabi oscillation) with the initial condition $\mathbf{r}_{I}(0) = (0,0,-1)$.
     }
     \label{Figure0}
\end{figure}

What happens if one does not neglect the effect of the rotation of $\mathbf{r}_{I}(t)$ around $\mathbf{\hat{w}}(2\omega_{1}t)$? We know that the rotation of $\mathbf{r}_{I}(t)$ induced by $\mathbf{\hat{w}}(2\omega_{1}t)$ approximately averages to zero during each complete turn of $\mathbf{\hat{w}}(2\omega_{1}t)$ around the $z$-axis. Hence, the effect on $\mathbf{r}_{I}(t)$ induced by $\mathbf{\hat{w}}(2\omega_{1}t)$ is going to be a small alteration of the  \textit{slow precessional-motion} around the $x$-axis. In fact, this small alteration \textit{looks like a fast nutational-motion} of the Bloch vector $\mathbf{r}_{I}(t)$ that involves the angular velocity $2\omega_{1}$ of $\mathbf{\hat{w}}(2\omega_{1}t)$, see Fig. \ref{Figure0a}. In the following we refer to the motion of the Bloch vector that \textit{looks like a fast nutational-motion} simply as the \textit{nutational-motion}. Therefore, the Bloch vector has an evolution similar to that of the motion of a symmetric top with one point fixed \cite{Marion}: the Bloch vector performs a \textit{nutational-precessional motion}, see Fig. \ref{Figure0a}.

Now assume that $\Omega_{0} \lesssim 2\omega_{1}$. In this case one can approximately separate the motion of the Bloch vector $\mathbf{r}_{I}(t)$ into a \textit{fast nutational motion} involving the angular frequency $2\omega_{1}$ and a \textit{slow precessional motion} around $x$-axis with angular velocity $-\Omega_{0}$ because $\mathbf{\hat{w}}(2\omega_{1}t)$ rotates around the $z$-axis faster than $\mathbf{r}_{I}(t)$ rotates around $\mathbf{\hat{w}}(2\omega_{1}t)$ and $\mathbf{\hat{x}}$. Nevertheless, the \textit{nutational-motion} is going to be larger for larger $\Omega_{0}$, see Fig. \ref{Figure0}. The origin of this is that, during one complete turn of $\mathbf{\hat{w}}(2\omega_{1}t)$ around the $z$-axis, $\mathbf{r}_{I}(t)$ changes appreciably and the rotation of $\mathbf{r}_{I}(t)$ induced by $\mathbf{\hat{w}}(2\omega_{1}t)$ does not average to zero:  using the notation above, the rotation of $\mathbf{r}_{I}(t)$ induced by $\mathbf{\hat{w}}(2\omega_{1}t)$  at any time $t = t_{2} \in [t_{1}, t_{1} +(T/2) )$  is only partially cancelled by the rotation induced by $\mathbf{\hat{w}}(2\omega_{1}t)$ at the later time $t = t_{2} + T/2 \in[t_{1}+(T/2), t_{1}+T)$. 

Finally, assume that $2\omega_{1} \lesssim \Omega_{0}$. In this case it appears that one cannot separate the motion of the Bloch vector $\mathbf{r}_{I}(t)$ into a \textit{nutational-precessional motion} since now $\mathbf{\hat{w}}(2\omega_{1}t)$ rotates around the $z$-axis at a slower angular velocity than that with which $\mathbf{r}_{I}(t)$ rotates around $\mathbf{\hat{w}}(2\omega_{1}t)$ and $\mathbf{\hat{x}}$, see Fig. \ref{Figure0b}. One then returns to the interpretation in the paragraph after equation (\ref{TE34fR}): the Bloch vector $\mathbf{r}_{I}(t)$ rotates around $\mathbf{\hat{Q}}(t)$ with an angular velocity $q(t)$. 

Now, we use the geometric description presented above to describe how the multiple-scales method works. We assume that
\begin{eqnarray}
\label{supuestoEscalas}
\Omega_{0} &\ll& 2\omega_{1} \ ,
\end{eqnarray}
so that we can clearly separate the motion of the Bloch vector into a \textit{slow precessional motion} around $\mathbf{\hat{x}}$ with angular velocity $-\Omega_{0}$ and a \textit{fast nutational motion} involving the angular velocity $2\omega_{1}$. The objective of the multiple-scales method is to separate the differential equation (\ref{TE34}) governing the motion of the Bloch vector into two coupled equations, one describing the \textit{fast nutational motion} and the other describing the \textit{slow precessional motion}. In order to do this, the first step is to identify a perturbation parameter. This is done by recalling from (\ref{supuestoEscalas}) that one has a fast angular speed $2\omega_{1}$ and a slow angular speed $\Omega_{0}$. Hence, the appropriate perturbation parameter is 
\begin{eqnarray}
\label{eps0}
\epsilon &=& \frac{\Omega_{0}}{2\omega_{1}} \  \ll  \ 1 \ .
\end{eqnarray}
Notice that $\epsilon$ is a positive, non dimensional quantity.

The second step is to define two new variables:
\begin{eqnarray}
\label{2E}
t_{1} &=& 2\omega_{1}t = \tau \quad \mbox{(the fast time-scale)}, \cr
t_{2} &=& \Omega_{0}t = \epsilon \tau\ \quad \mbox{(the slow time-scale)}.
\end{eqnarray}
Notice that we have expressed $t_{1}$ and $t_{2}$ as functions of a non dimensional time $\tau$ so that the perturbation parameter $\epsilon$ appears explicitly. Moreover, from the discussion in the preceding paragraphs observe that $t_{1}$ is a (nondimensional) time variable associated with the \textit{fast nutational motion} of the Bloch vector, while $t_{2}$ is a (nondimensional) time variable associated with the \textit{slow precessional motion} of the Bloch vector.

The third step is to define a new Bloch vector that incorporates the two new (nondimensional) time variables:
\begin{eqnarray}
\label{NewZ}
\mathbf{K}\left[ t_{1}(\tau), t_{2}(\tau) \right] &=& \mathbf{r}_{I}\left( \frac{\tau}{2\omega_{1}} \right) \ .
\end{eqnarray}

The fourth step is to obtain a partial differential equation in the variables $t_{1}$ and $t_{2}$ for $\mathbf{K}(t_{1},t_{2})$. This is done by substituting (\ref{NewZ}) into (\ref{TE34}). A straightforward calculation leads to the equation
\begin{eqnarray}
\label{TE6}
\frac{\partial \mathbf{K}}{\partial t_{1}} (t_{1},t_{2}) &=& -\epsilon \left[ \mathbf{\hat{x}} + \mathbf{\hat{w}}(t_{1}) \right] \times \mathbf{K}(t_{1}, t_{2}) - \epsilon \frac{\partial \mathbf{K}}{\partial t_{2}} (t_{1},t_{2}) \  . \cr
&&
\end{eqnarray} 
This indicates that the new Bloch vector $\mathbf{K}(t_{1},t_{2})$ rotates around the unit vector $\left[ \mathbf{\hat{x}} + \mathbf{\hat{w}}(t_{1}) \right] / \left\vert \mathbf{\hat{x}} + \mathbf{\hat{w}}(t_{1}) \right\vert $ with an angular velocity $-\epsilon \vert \mathbf{\hat{x}} + \mathbf{\hat{w}}(t_{1}) \vert$ and that this rotation is altered by the term $-\epsilon ( \partial \mathbf{K}/\partial t_{2}) (t_{1},t_{2})$. Since this last quantity involves a partial derivative with respect to the \textit{precessional time} $t_{2}$, it could be interpreted as the \textit{precessional velocity} of the new Bloch vector.  

The final step is to solve (\ref{TE6}). This is done by assuming that $\mathbf{K}(t_{1},t_{2})$ has an asymptotic expansion of the form
\begin{eqnarray}
\label{TE7}
\mathbf{K} &\sim& \mathbf{K}_{0} + \epsilon \mathbf{K}_{1} + \epsilon^{2}\mathbf{K}_{2} + ... \ .
\end{eqnarray}
Then, one substitutes this asymptotic expansion into the differential equation for $\mathbf{K}(t_{1},t_{2})$ given in  (\ref{TE6}) and into the initial condition $\mathbf{K}(0,0) = \mathbf{r}_{I}(0)$. Afterwards, one equates equal powers of $\epsilon$ and one arrives at the following first three initial value problems:
\begin{eqnarray}
\label{TE8}
&\mathcal{O}(1):&   \frac{\partial \mathbf{K}_{0}}{\partial t_{1}} (t_{1},t_{2}) \ = \ 0 \ , \cr 
&& \mathbf{K}_{0}(0,0) = \mathbf{r}_{I}(0) \ . \cr
&& \cr
&\mathcal{O}(\epsilon):&   \frac{\partial \mathbf{K}_{1}}{\partial t_{1}} (t_{1},t_{2}) \ = \ - \left[ \mathbf{\hat{x}} + \mathbf{\hat{w}}(t_{1}) \right] \times \mathbf{K}_{0}(t_{1}, t_{2}) \cr
&& \quad \quad \quad \quad \quad \quad \quad - \frac{\partial \mathbf{K}_{0}}{\partial t_{2}} (t_{1},t_{2}) \ , \cr
&& \mathbf{K}_{1}(0,0) = \mathbf{0} \ . \cr
&& \cr
&\mathcal{O}(\epsilon^{2}):&   \frac{\partial \mathbf{K}_{2}}{\partial t_{1}} (t_{1},t_{2}) \ = \ - \left[ \mathbf{\hat{x}} + \mathbf{\hat{w}}(t_{1}) \right] \times \mathbf{K}_{1}(t_{1}, t_{2}) \cr
&& \quad \quad \quad \quad \quad \quad \quad - \frac{\partial \mathbf{K}_{1}}{\partial t_{2}} (t_{1},t_{2}) \ , \cr
&& \mathbf{K}_{2}(0,0) = \mathbf{0} \ . 
\end{eqnarray}
All of these initial value problems can be solved exactly and all the details are provided in the appendix. In this section we limit ourselves to describing the process and we provide an interpretation of the equations.

From (\ref{TE8}) one finds that the solution of the $\mathcal{O}(1)$ differential equation is 
\begin{eqnarray}
\label{TE9}
\mathbf{K}_{0}(t_{1},t_{2}) &=& \mathbf{K}_{0}(0,t_{2}) \ .
\end{eqnarray}
Therefore, $\mathbf{K}_{0}(t_{1},t_{2})$ only depends on the \textit{slow precessional time} $t_{2}$ and not on the \textit{fast nutational time} $t_{1}$. In other words, $\mathbf{K}_{0}(t_{1},t_{2})$ only describes the \textit{slow precessional motion} of the Bloch vector $\mathbf{r}_{I}(t)$. 

One then substitutes (\ref{TE9}) in the $\mathcal{O}(\epsilon)$ differential equation in (\ref{TE8}). Solving the resulting equation one finds that one must eliminate secular terms from $\mathbf{K}_{1}(t_{1},t_{2})$, that is, terms that become unbounded as $t_{1} \rightarrow + \infty$ and that destroy the order of the asymptotic expansion in (\ref{TE7}). The secular terms are eliminated if $\mathbf{K}_{0}(t_{1},t_{2})$ satisfies the following differential equation:
\begin{eqnarray}
\label{TE10}
\frac{\partial \mathbf{K}_{0}}{\partial t_{2}} (0,t_{2}) &=& - \mathbf{\hat{x}} \times \mathbf{K}_{0}(0,t_{2}) \ .
\end{eqnarray}
The first thing to observe is that (\ref{TE10}) is equivalent to the equation for the Bloch vector $\mathbf{r}_{I}(t)$ in the RWA given in (\ref{TE34rwa}): if one takes the partial derivative with respect to $t_{2}$ of $\mathbf{K}_{0}(0,t_{2}) = \mathbf{r}_{I}(t_{2}/\Omega_{0})$ and one uses (\ref{TE34rwa}), then one obtains (\ref{TE10}). Since $\mathbf{K}_{0}(t_{1},t_{2})$ does not depend on $t_{1}$, see (\ref{TE9}), it follows that  $\mathbf{K}_{0}(t_{1},t_{2})$ coincides with the Bloch vector in the RWA approximation.

Substituting the secular equation in (\ref{TE10}) and the result (\ref{TE9}) into the $\mathcal{O}(\epsilon)$ differential equation, one finds that $\mathbf{K}_{1}(t_{1},t_{2})$ must satisfy the following equation:
\begin{eqnarray}
\label{TE9b}
 \frac{\partial \mathbf{K}_{1}}{\partial t_{1}} (t_{1},t_{2}) \ = \ - \mathbf{\hat{w}}(t_{1}) \times \mathbf{K}_{0}(0, t_{2}) \ . 
\end{eqnarray}
From (\ref{TE9b}) one finds that the origin of the $t_{1}$-dependence of $\mathbf{K}_{1}(t_{1},t_{2})$ is given by $\mathbf{\hat{w}}(t_{1})$. Notice that $(\partial \mathbf{K}_{1} / \partial t_{1})(t_{1},t_{2})$ can be interpreted as the \textit{nutational velocity} of  $\mathbf{K}_{1} (t_{1},t_{2})$, since $t_{1}$ is the (nondimensional) \textit{nutational time}. Therefore, (\ref{TE9b}) indicates that the \textit{nutational velocity} of  $\mathbf{K}_{1} (t_{1},t_{2})$ is determined by $\mathbf{\hat{w}}(t_{1})$.

After solving (\ref{TE9b}) one substitutes the result in the $\mathcal{O}(\epsilon^{2})$ differential equation for $\mathbf{K}_{2}(t_{1},t_{2})$. Solving the resulting equation one finds that one must also eliminate secular terms from $\mathbf{K}_{2}(t_{1},t_{2})$. These are eliminated if $\mathbf{K}_{1}(t_{1},t_{2})$ satisfies the following equation:
\begin{eqnarray}
\label{TE13}
\frac{\partial \mathbf{K}_{1}}{\partial t_{2}} (0,t_{2}) &=& - \mathbf{\hat{x}} \times \mathbf{K_{1}} (0,t_{2})  - \frac{1}{2} \mathbf{\hat{z}} \times \mathbf{K}_{0}(0,t_{2}) \ .
\end{eqnarray}
Since $t_{2}$ is the (non dimensional) \textit{slow precessional time}, it follows that $ \mathbf{K}_{1} (0,t_{2})$ slowly rotates around $\mathbf{\hat{x}}$ with a (nondimensional) angular velocity equal to $-1$ and that this motion is altered by the second term on the righthand side of (\ref{TE13}).

Up to this point  $\mathbf{K}_{0} (t_{1},t_{2})$ and $\mathbf{K}_{1} (t_{1},t_{2})$ have both been determined so one can construct a one- or two-term approximation of $\mathbf{r}_{I}(t)$ using the  the definition of $\mathbf{K}(t_{1},t_{2})$ in terms of $\mathbf{r}_{I}(t)$ given in (\ref{NewZ}) and the asymptotic expansion of $\mathbf{K}(t_{1},t_{2})$ in (\ref{TE7}). The one-term approximation is given by 
\begin{eqnarray}
\label{Nuevo1termino}
\mathbf{r}_{I}(t) &\simeq& \mathbf{K}_{0} (2\omega_{1}t, \Omega_{0}t) \ ,
\end{eqnarray}
while the two-term approximation is given by 
\begin{eqnarray}
\label{Nuevo2termino}
\mathbf{r}_{I}(t) &\simeq& \mathbf{K}_{0} (2\omega_{1}t, \Omega_{0}t) + \epsilon \mathbf{K}_{1} (2\omega_{1}t, \Omega_{0}t) \ .
\end{eqnarray}
From the discussion in the preceding paragraphs, the one-term approximation is identical to the solution in the RWA and, as such, describes only the precessional motion of the Bloch vector. On the other hand, the two-term approximation is more accurate and describes the \textit{nutational-precesional motion} of the Bloch vector to good approximation. In the next sections we give an explicit expression of the two term approximation and we determine its accuracy.

%%%%%%%%%%%%%%%%%%%%%%%%%%%%%%%%%%
%%%%%%%%%%%%%%%%%%%%%%%%%%%%%%%%%%
%%%%%%%%%%%%%%%%%%%%%%%%%%%%%%%%%%
%%%%%%%%%%%%%%%%%%%%%%%%%%%%%%%%%%

\subsection{The two-term approximation}

The method of multiple scales using the two time-scales in (\ref{2E}) leads to the following two-term approximation:
\begin{widetext}
\begin{eqnarray}
\label{Aprox2terminoU}
\rho_{12}(t) &=& 
\rho_{12}(0)\left[ \mbox{cos}^{2}\left( \frac{\Omega_{0}t}{2} \right) - i\frac{\epsilon}{2}e^{-i2\omega_{1}t}\mbox{sin}(\Omega_{0}t) \right] + \rho_{12}(0)^{*}\left[ \mbox{sin}^{2}\left( \frac{\Omega_{0}t}{2} \right) -i\frac{\epsilon}{2}\mbox{sin}(\Omega_{0}t) \left( 1 - e^{-i2\omega_{1}t} \right) \right]  \cr
&& + \alpha_{30}(0) \frac{i}{2} \left\{   \mbox{sin}\left( \Omega_{0}t \right) -i\epsilon \left[ \mbox{cos}^{2}\left( \frac{\Omega_{0}t}{2} \right) - e^{-i2\omega_{1}t}\mbox{cos}(\Omega_{0}t)  \right] \right\} , \cr
&& \cr
&& \cr
\alpha_{30}(t) &=&
 i \rho_{12}(0) \left( \  \mbox{sin}\left( \Omega_{0}t \right)   +i\frac{\epsilon}{2} \left\{ 1 -2\mbox{cos}(2\omega_{1}t) + \mbox{cos}(\Omega_{0}t) \Big[ 1 -i2\mbox{sin}(2\omega_{1}t) \Big]  \right\} \right) \cr
&& - i \rho_{12}(0)^{*} \left( \  \mbox{sin}\left( \Omega_{0}t \right)   -i\frac{\epsilon}{2} \Big\{ 1 -2\mbox{cos}(2\omega_{1}t) + \mbox{cos}(\Omega_{0}t) \left[ 1 +i2\mbox{sin}(2\omega_{1}t) \right]  \Big\} \right) \ , \cr
&& + \alpha_{30}(0)\left[ \ \mbox{cos}\left(\Omega_{0}t\right) - \epsilon  \mbox{sin}(\Omega_{0}t)\mbox{sin}(2\omega_{1}t) \ \right]  .
\end{eqnarray}
\end{widetext}
Notice that $\alpha_{30}(t)$ is indeed a real quantity and that the two-term approximation (\ref{Aprox2terminoU}) is not much more complicated than the solution in the RWA approximation given in (\ref{RWAsolucionE}), since the new terms are those that are multiplied by $\epsilon$.

In order to give a geometrical interpretation of (\ref{Aprox2terminoU}) it is convenient to write the result in terms of the Bloch vector. A straightforward calculation using the definition of the Bloch vector given in (\ref{Matriz3})-(\ref{ComponentesBloch}) and the two-term approximation given in (\ref{Aprox2terminoU}) leads to the following very simple expression of the two-term approximation of the Bloch vector: 
\begin{eqnarray}
\label{TE43}
\mathbf{r}_{I}(t) &=& \mathbf{r}_{I}^{\mbox{\tiny RWA}}(t) - \epsilon  \mathbf{w}_{3}(t) \times  \mathbf{r}_{I}^{\mbox{\tiny RWA}}(t) \ ,
\end{eqnarray}
where $\mathbf{r}_{I}^{\mbox{\tiny RWA}}(t)$ is the Bloch vector in the RWA given in (\ref{TE36Bloch}) and
\begin{eqnarray}
\label{TE42}
\mathbf{\hat{w}}_{2}(t) &=& \mathbf{\hat{y}} \mbox{sin}\left( \frac{\Omega_{0}t}{2} \right) +\mathbf{\hat{z}} \mbox{cos}\left( \frac{\Omega_{0}t}{2} \right) \ , \cr
&& \cr
\mathbf{w}_{3}(t) &=& \int_{0}^{2\omega_{1}t} d\tau' \mathbf{\hat{w}}(\tau') + \mbox{sin}\left( \frac{\Omega_{0}t}{2} \right) \mathbf{\hat{w}}_{2}(t) \ , \cr
&& \cr
&=& \left[ \mathbf{\hat{x}}\mbox{sin}(2\omega_{1}t) + \mathbf{\hat{y}}\mbox{cos}(2\omega_{1}t) \right] \ \cr
&& + \mbox{cos}\left( \frac{\Omega_{0}t}{2} \right)\left[ \mathbf{\hat{z}}\mbox{sin}\left( \frac{\Omega_{0}t}{2} \right) - \mathbf{\hat{y}}\mbox{cos}\left( \frac{\Omega_{0}t}{2} \right)  \right] \ . \cr
&&
\end{eqnarray}

We now discuss the effects of the counterrotating terms in the evolution of the system. These are embodied by the corrections to the solution in the RWA given in (\ref{RWAsolucionE}). Notice that the difference between the two solutions in (\ref{RWAsolucionE}) and (\ref{Aprox2terminoU}) consists of the terms multiplied by $\epsilon$. Hence, the effect of the counterrotating terms is to introduce terms that are multiplied by $\epsilon$ and that oscillate at the angular frequencies $2\omega_{1}$ and $\Omega_{0}$. 

The corrections can be interpreted easily by using the geometric picture introduced in the previous section. Recall that it was established that the the Bloch vector $\mathbf{r}_{I}(t)$ presents a \textit{nutational-precessional motion} when $0< \epsilon < 1$, which is the case we are considering. It was also deduced that in the RWA one neglects the \textit{fast nutational motion} and one only keeps the \textit{slow-precessional motion}, so that $\mathbf{r}_{I}(t)$ only rotates around the $x$-axis with an angular velocity $-\Omega_{0}$. In addition, it was presented that the counterrotating terms are responsible for the \textit{fast nutational motion} of $\mathbf{r}_{I}(t)$. With this in mind we proceed to interpret (\ref{TE43}). First, the two-term approximation of the Bloch vector given in (\ref{TE43}) is composed of two terms. The first one is $\mathbf{r}_{I}^{\mbox{\tiny RWA}}(t)$, so $\mathbf{r}_{I}(t)$ in (\ref{TE43}) preserves the \textit{slow-precessional motion} around the $x$-axis described by the RWA. The second term is smaller because it is proportional to $\epsilon$. Moreover, it is perpendicular to $\mathbf{r}_{I}^{\mbox{\tiny RWA}}(t)$ and depends on the angular frequency $2\omega_{1}$ which is associated with the counterrotating terms. Therefore, the second term describes the \textit{fast nutational motion} of the Bloch vector.

Now, we interpret the vector $\mathbf{w}_{3}(t)$ included in the second term on the righthand side of (\ref{TE43}). From (\ref{TE42}) observe that $\mathbf{w}_{3}(t)$ is composed of two parts: a vector $\left[ \mathbf{\hat{x}}\mbox{sin}(2\omega_{1}t) + \mathbf{\hat{y}}\mbox{cos}(2\omega_{1}t) \right]$ contained in the $xy$-plane that is rotating around the $z$-axis with a \textit{fast} angular velocity $2\omega_{1}$ and a vector $\mbox{cos}\left( \Omega_{0}t/2 \right)\left[ \mathbf{\hat{z}}\mbox{sin}\left( \Omega_{0}t/2 \right) -\mathbf{\hat{y}}\mbox{cos}\left( \Omega_{0}t/2 \right)  \right]$  contained in the $yz$-plane that is rotating around the $x$-axis with a \textit{slow} angular velocity $-\Omega_{0}$ (recall that a positive (negative) angular velocity indicates that the rotation is in the counterclockwise (clockwise) sense). Therefore, the \textit{nutational motion} of the Bloch vector can be divided into a \textit{fast nutational motion} involving the angular velocity $2\omega_{1}$ and a \textit{slow nutational motion} involving the angular velocity $-\Omega_{0}$. The fast nutational motion is due to  $\left[ \mathbf{\hat{x}}\mbox{sin}(2\omega_{1}t) + \mathbf{\hat{y}}\mbox{cos}(2\omega_{1}t) \right]$, while the slow nutational motion is due to $\mbox{cos}\left( \Omega_{0}t/2 \right)\left[ \mathbf{\hat{z}}\mbox{sin}\left( \Omega_{0}t/2 \right) -\mathbf{\hat{y}}\mbox{cos}\left( \Omega_{0}t/2 \right)  \right]$. In addition, observe from (\ref{TE42}) that $\left[ \mathbf{\hat{x}}\mbox{sin}(2\omega_{1}t) + \mathbf{\hat{y}}\mbox{cos}(2\omega_{1}t) \right]$ arises from the integral of $\mathbf{\hat{w}}(\tau)$, the vector associated with the counterrotating terms in the equation of motion of the Bloch vector in (\ref{TE34}). Finally, from (\ref{TE43}) and (\ref{TE42}) notice that the dependence of $\mathbf{r}_{I}(t)$ on the integral of $\mathbf{\hat{w}}(\tau)$ contained in $\mathbf{w}_{3}(t)$ arises when one solves (\ref{TE9b}) in the multiple scales method.

Also, the two-term approximation allows one to have a quantitative criterion that indicates when the solution in the RWA is accurate. If one subtracts $\rho_{12}^{\mbox{\tiny RWA}}(t)$ and $\alpha_{30}^{\mbox{\tiny RWA}}(t)$ respectively from $\rho_{12}(t)$ and $\alpha_{30}(t)$ given in (\ref{Aprox2terminoU}) and takes the absolute value, one gets
\begin{widetext}
\begin{eqnarray}
\label{criterio1}
\left\vert \rho_{12}(t) - \rho_{12}^{\mbox{\tiny RWA}}(t) \right\vert &=& \Big\vert \alpha_{30}(0) \frac{\epsilon}{2}  \left[ \mbox{cos}^{2}\left( \frac{\Omega_{0}t}{2} \right) - e^{-i2\omega_{1}t}\mbox{cos}(\Omega_{0}t)  \right]  \cr
&& +\rho_{12}(0)\left[ - i\frac{\epsilon}{2}e^{-i2\omega_{1}t}\mbox{sin}(\Omega_{0}t) \right] + \rho_{12}(0)^{*}\left[ -i\frac{\epsilon}{2}\mbox{sin}(\Omega_{0}t) \left( 1 - e^{-i2\omega_{1}t} \right) \right] \Big\vert , \cr
&\leq& 2\epsilon \ , \cr
&& \cr
&& \cr
\left\vert \alpha_{30}(t) - \alpha_{30}^{\mbox{\tiny RWA}}(t) \right\vert &=& \Big\vert \alpha_{30}(0) (- \epsilon ) \mbox{sin}(\Omega_{0}t)\mbox{sin}(2\omega_{1}t)  -\rho_{12}(0)\frac{\epsilon}{2} \left\{ 1 -2\mbox{cos}(2\omega_{1}t) + \mbox{cos}(\Omega_{0}t) \Big[ 1 -i2\mbox{sin}(2\omega_{1}t) \Big]  \right\}  \cr
&& - \rho_{12}(0)^{*} \frac{\epsilon}{2} \Big\{ 1 -2\mbox{cos}(2\omega_{1}t) + \mbox{cos}(\Omega_{0}t) \left[ 1 +i2\mbox{sin}(2\omega_{1}t) \right]  \Big\}   \Big\vert \ , \cr
&\leq& 4\epsilon \ ,
\end{eqnarray}
\end{widetext}
so that
\begin{eqnarray}
\label{criterio}
\left\vert \rho_{12}(t) - \rho_{12}^{(\mbox{\tiny RWA})}(t) \right\vert ,  \left\vert \alpha_{30}(t) - \alpha_{30}^{(\mbox{\tiny RWA})}(t) \right\vert &\leq& 4\epsilon  = \frac{2\Omega_{0}}{\omega_{1}}\ . \cr
&&
\end{eqnarray}
We note that, to obtain the bounds in (\ref{criterio1}), we used the triangle inequality and that $\vert \rho_{12}(0) \vert \leq 1/2$ and $\vert \alpha_{30}(0) \vert \leq 1$. The two latter inequalities hold because $\rho_{I}(0)$ is a density operator and its eigenvalues $(1 \pm \sqrt{\alpha_{30}(0)^{2} + 4 \vert \rho_{12}(0) \vert^{2}})/2$ must be non negative.

From (\ref{criterio}) it follows that the corrections introduced to the solution in the RWA are $\leq 4\epsilon$. We illustrate (\ref{criterio}) with the parameters from the cavity QED experiments in \cite{Maser2} where the RWA is known to hold (although a quantum field instead of a classical one appears in those experiments). Those experiments have $\omega_{1} = 2\pi \times 51.1 \times 10^{9}$ 1/s and $\Omega_{0} = 2\pi \times 47 \times 10^{3}$ 1/s, so that $\epsilon = 5 \times 10^{-7}$ and the corrections to the solution in the RWA are $\leq 4\epsilon = 2 \times 10^{-6}$.

%%%%%%%%%%%%%%%%%%%%%%%%%%%%%%%%%%
%%%%%%%%%%%%%%%%%%%%%%%%%%%%%%%%%%
%%%%%%%%%%%%%%%%%%%%%%%%%%%%%%%%%%
%%%%%%%%%%%%%%%%%%%%%%%%%%%%%%%%%% 
\subsection{Density operator defined by the two-term approximation}

When one substitutes the solution in the RWA given in (\ref{RWAsolucionE}) into the matrix (\ref{Matriz}), one obtains a density matrix, since (\ref{RWAsolucionE}) was obtained by solving von Neumann's equation (\ref{13}) in the IP with the approximate interaction Hamiltonian in (\ref{15b}). One also obtains a density matrix with the one-term approximation because it is identical to the solution in the RWA. However, the same cannot be said when one substitutes the two-term approximation given in (\ref{Aprox2terminoU}) into the matrix (\ref{Matriz}). In some cases one obtains a density matrix and in others one does not. For example, if $\rho_{I}(0)$ represents a pure state, then one does not obtain a density matrix. The reason for this is that the length of the Bloch vector associated with the two-term approximation is slightly larger than the length of the initial Bloch vector (the details are given in Appendix A.4). A way to remedy this is explained below and it is similar to normalizing a vector.

Assume that the initial density operator $\rho_{I}(0) = \rho (0) $ is a pure state (recall that the Sch\"{o}dinger picture and the IP coincide at $t=0$, see (\ref{11}) and (\ref{12})). First calculate the values of $\rho_{12}(t)$ and $\alpha_{30}(t)$ using the two-term approximation given in (\ref{Aprox2terminoU}). Then construct the matrix 
\begin{eqnarray}
\label{Paso2}
[\rho_{I}^{(p)}(t) ]_{\gamma} = \left(
\begin{array}{cc}
1 - \rho_{22}^{(p)}(t) & \frac{\rho_{12}(t)}{N(t)} \cr
\frac{\rho_{12}(t)^{*}}{N(t)} & \rho_{22}^{(p)}(t)
\end{array}
\right)
\end{eqnarray}
where
\begin{eqnarray}
\label{Paso3b}
\rho_{22}^{(p)} (t) &=& \frac{\alpha_{30}(t)}{2N(t)} + \frac{1}{2} \   , \quad N(t) \ = \ \sqrt{ \alpha_{30}(t)^{2} + 4 \vert \rho_{12}(t) \vert^{2}} \   . \cr
&&
\end{eqnarray}
By construction $[\rho_{I}^{(p)}(t) ]_{\gamma}$ is Hermitian and has trace equal to $1$ (recall that $\alpha_{30}(t)$ is a real quantity). Moreover, it is straightforward to show that its eigenvalues are $1$ and $0$. Hence, $[\rho_{I}^{(p)}(t) ]_{\gamma}$ is the density matrix of a pure state and it represents the approximate state of the system at time $t$ in the IP. 

It is easy to give a geometrical interpretation to what is being done in (\ref{Paso2}). Given the values of $\rho_{12}(t)$ and $\alpha_{30}(t)$ in the two-term approximation in (\ref{Aprox2terminoU}), one can construct the approximate Bloch vector $\mathbf{r}_{I}(t)$ in (\ref{TE43}). Since $\rho_{I}(0) = \rho (0) $ is a pure state, the (Euclidean) norm of $\mathbf{r}_{I}(t)$ should always be one. Nevertheless, it happens that the (Euclidean) norm of $\mathbf{r}_{I}(t)$ is greater than or equal to $1$ (see Appendix A.4). Therefore, one should simply normalize $\mathbf{r}_{I}(t)$ and take $\mathbf{r}_{I}(t)/\vert \mathbf{r}_{I}(t) \vert$ to be the approximate Bloch vector of the system in the IP. It is straightforward to show using (\ref{Matriz3})-(\ref{ComponentesBloch}) that the Bloch vector associated with the density matrix $[\rho_{I}^{(p)}(t) ]_{\gamma}$ given in (\ref{Paso2}) is precisely $\mathbf{r}_{I}(t)/\vert \mathbf{r}_{I}(t) \vert$.
 
Now assume that the initial density operator $\rho_{I}(0) = \rho (0) $ is a mixed state. First express $\rho_{I}(0) = \rho (0) $ as a convex combination of pure states:
\begin{eqnarray}
\label{Paso1}
\rho_{I}(0) = \rho (0) = \sum_{k=1}^{n} a_{k}\rho_{k}(0) \ ,
\end{eqnarray}
where $\rho_{k}(0)$ is a density operator representing a pure state, $a_{k}\in [0,1]$, and $\sum_{k=1}^{n} a_{k} = 1$. Then, for each $k$ calculate the pure-state density operator $\rho_{kI}^{(p)}(t)$ associated with $\rho_{k}(0)$ using (\ref{Aprox2terminoU}), (\ref{Paso2}), and (\ref{Paso3b}). It follows that the density operator of the system in the IP is
\begin{eqnarray}
\label{Paso1b}
\rho_{I}(t) =  \sum_{k=1}^{n} a_{k}\rho_{kI}^{(p)}(t) \ .
\end{eqnarray}

%%%%%%%%%%%%%%%%%%%%%%%%%%%%%%%%%%
%%%%%%%%%%%%%%%%%%%%%%%%%%%%%%%%%%
%%%%%%%%%%%%%%%%%%%%%%%%%%%%%%%%%%
%%%%%%%%%%%%%%%%%%%%%%%%%%%%%%%%%%

\subsection{Accuracy of the two-term approximation}

The theory of multiple scales \cite{Holmes} tells us that the approximate solutions using the two time-scales in (\ref{2E}) hold at least for a time interval of the form
\begin{eqnarray}
\label{Intervalo2E}
 0 \leq t_{2} = \Omega_{0}t  \leq \mathcal{O}\left( 1 \right)  \quad \Leftrightarrow \quad 0 \leq t \leq \mathcal{O}\left( \frac{1}{\Omega_{0}}  \right) \ . 
\end{eqnarray}
Here and in the following $\mathcal{O}$ denotes the \textit{Big Oh} \cite{Holmes}. Recall from Sec. III that $2\pi /\Omega_{0}$ is the period for one Rabi oscillation, so that (\ref{Intervalo2E}) indicates that the approximate solutions obtained are accurate at least for $k$ times one Rabi oscillation with $k>0$.

We now determine explicitly the accuracy of the two-term approximation given in (\ref{Aprox2terminoU}) by comparing it to the numerical solution of von Neumann's equation (\ref{13}) with the complete interaction Hamiltonian in (\ref{15}). 

For each value of $\epsilon$ between $\epsilon_{min}$ and $\epsilon_{max}$ we calculated the following quantities (see Appendix A.5 for the details):
\begin{enumerate}
\item $E_{R}(\epsilon)$: it is the maximum relative error that one can have if one uses the two-term approximation in (\ref{Aprox2terminoU}) from $t = 0$ to $t = t_{max}$ with any initial condition that corresponds to a pure state; 
\item $E_{RN}(\epsilon)$: it is the maximum relative error that one can have if one uses the normalized two-term approximation in (\ref{Aprox2terminoU}), (\ref{Paso2}), and (\ref{Paso3b}) from $t = 0$ to $t = t_{max}$ with any initial condition that corresponds to a pure state; 
\item $E_{R}^{RWA}(\epsilon)$: it is the maximum relative error that one can have if one uses the approximation in the RWA given in (\ref{RWAsolucionE}) from $t = 0$ to $t = t_{max}$ with any initial condition that corresponds to a pure state. 
\end{enumerate}
In each case the relative error  is calculated as the norm of the corresponding approximate Bloch vector minus the exact (numerical) Bloch vector over the norm of the exact (numerical) Bloch vector (which is $1$ because the initial condition is a pure state).

\begin{figure}[htbp]
     \centering
     \subfloat[]{\label{Figure1a}\includegraphics[scale=0.45]{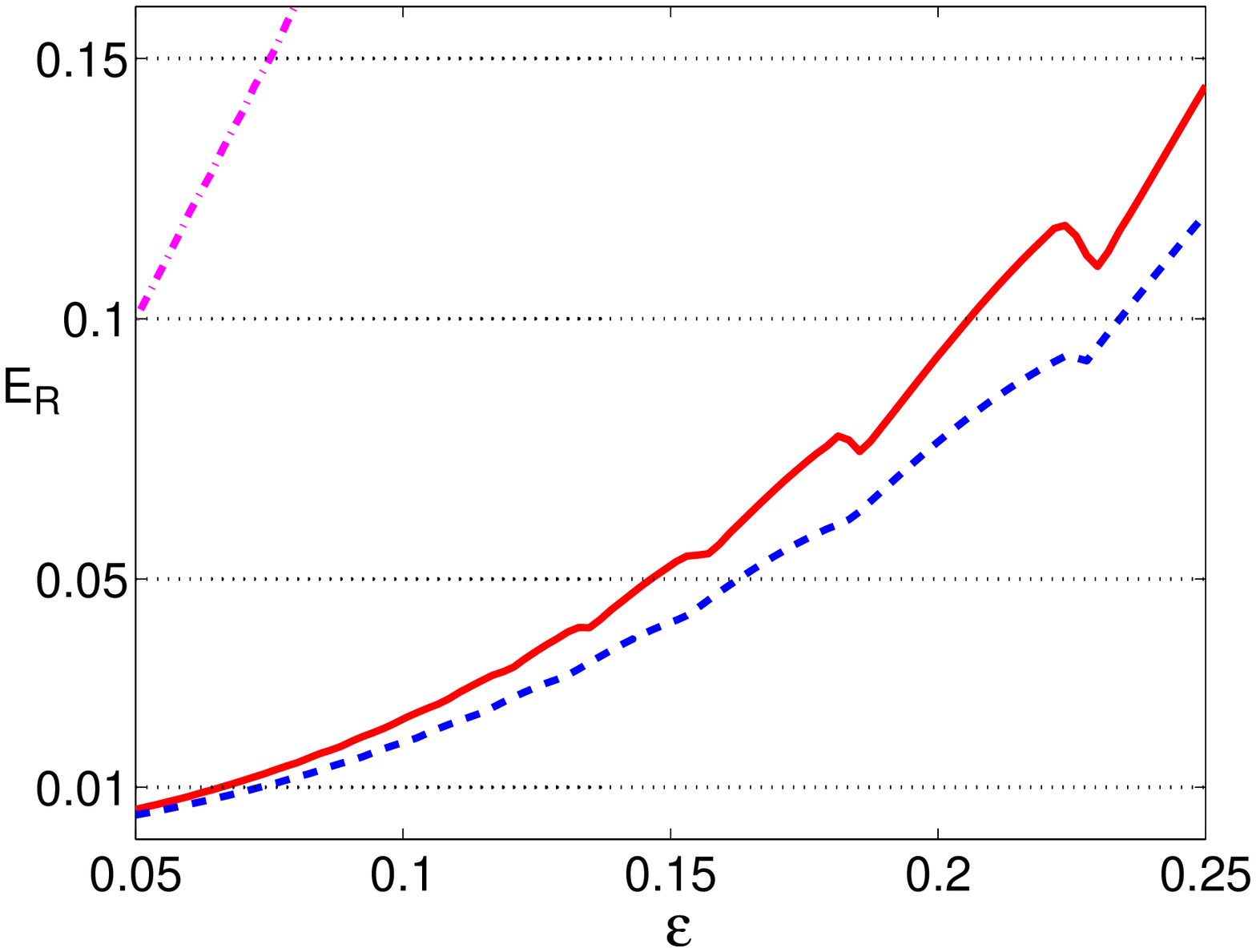}}\\
     \subfloat[]{\label{Figure1b}\includegraphics[scale=0.45]{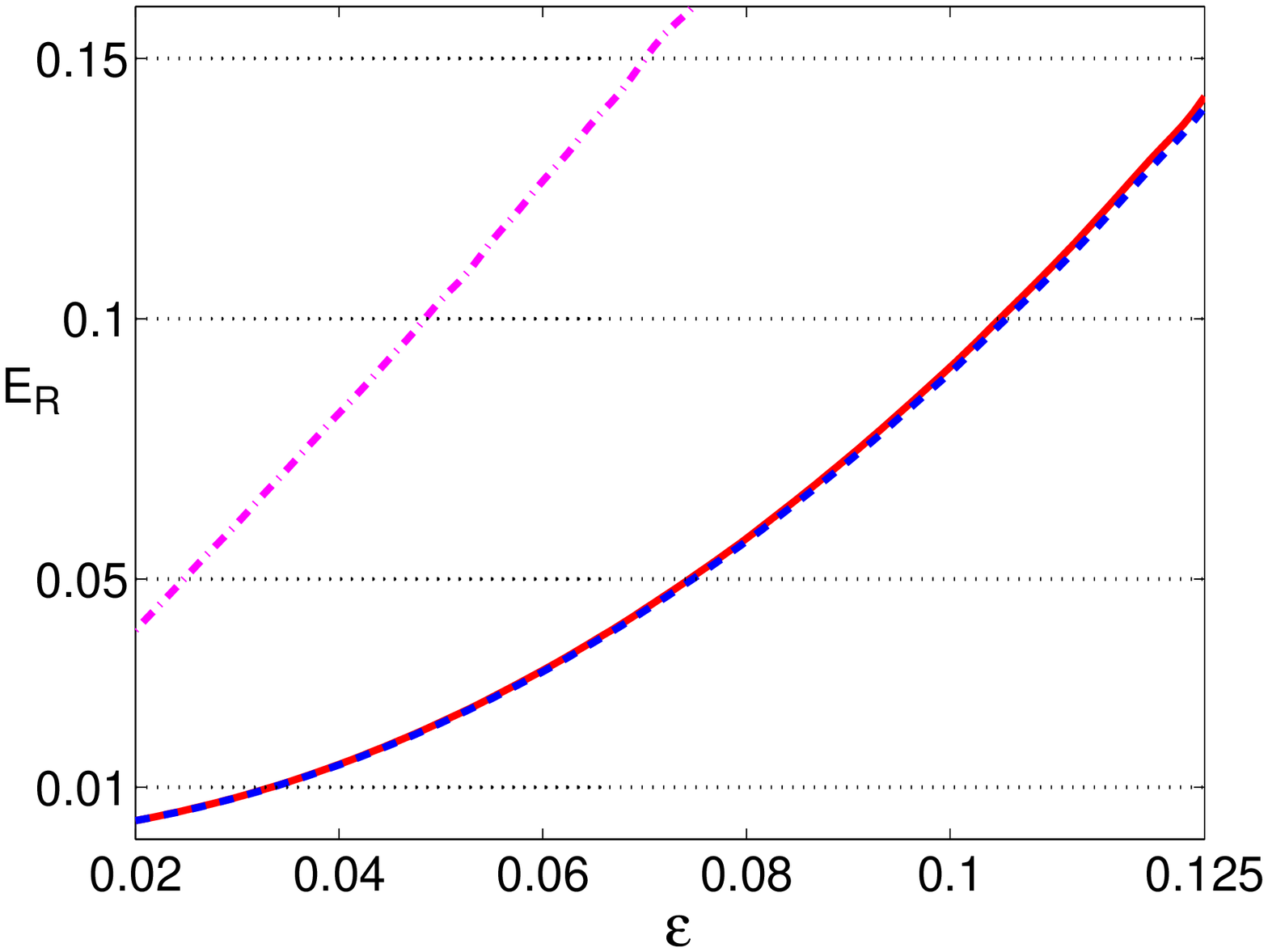}}
     \caption{(Color online) The figures show the maximum relative error $E_{R}$ of the approximate solutions as a function of $\epsilon$ given in (\ref{eps0}) for any pure-state initial condition. The maximum relative error $E_{R}(\epsilon)$ of the two-term approximation is shown in red-solid lines, while the maximum relative error $E_{RN}(\epsilon)$ of the two-term normalized approximation and the maximum relative error $E_{R}^{RWA}(\epsilon)$ of the solution in the RWA are shown in blue-dashed and magenta-dot-dashed lines, respectively. Horizontal-black-dotted lines indicate a relative error of  $15\%$, $10\%$, $5\%$, and $1\%$. Figure \ref{Figure1a} only considers the first Rabi oscillation and $1/20 \leq \epsilon \leq 1/4$, while Figure \ref{Figure1b} considers the first $10$ Rabi oscillations and $1/50 \leq \epsilon \leq 1/8$.}
     \label{Figure1}
\end{figure}

Figure \ref{Figure1}  shows the graphs of  $E_{R}(\epsilon)$ (red-solid line),  $E_{RN}(\epsilon)$ (blue-dashed line), and  $E_{R}^{RWA}(\epsilon)$ (magenta-dot-dashed line) as a function of $\epsilon$. Figure \ref{Figure1a} illustrates the results for $\epsilon_{min} = 0.05 = 1/20$, $\epsilon_{max} = 0.25 = 1/4$, and $t_{max}=2\pi/\Omega_{0}$ ($1$ Rabi oscillation), while Figure \ref{Figure1b} shows the results for $\epsilon_{min} = 0.02 = 1/50$, $\epsilon_{max} = 0.125 = 1/8$, and $t_{max}=10(2\pi/\Omega_{0})$ ($10$ Rabi oscillations). 

In Figure \ref{Figure1a} notice that both two-term approximations (normalized and non normalized)  are quite good, since the relative error is less than $15\%$ if $\epsilon \leq 1/4$ or, equivalently, if the Rabi frequency $\Omega_{0}$ is at most half the angular frequency of the field $\omega_{1}$. Moreover, the relative error is less than $1\%$ as soon as $\epsilon \leq 0.07$ ($\epsilon \leq 0.066$) for the two-term normalized (non normalized) solution. For comparison, the solution in the RWA is quite bad, since $E_{R}^{RWA}(\epsilon)$ has a relative error less than $15\%$ only when $\epsilon \leq 0.075$. Therefore, we conclude that both the two-term approximate solution given in (\ref{Aprox2terminoU}) and the two-term normalized approximate solution given in (\ref{Aprox2terminoU}) and (\ref{Paso2})-(\ref{Paso1b}) are accurate descriptions of the system from $t= 0$ to  $t = 2\pi/\Omega_{0}$, that is, during the first Rabi oscillation as long as $\Omega_{0} \leq \omega_{1}/2$.

In Figure \ref{Figure1b} observe that both two-term approximations (normalized and non normalized)  are also quite good, since the relative error is less than $15\%$ if $\epsilon \leq 1/8$ or, equivalently, if the Rabi frequency $\Omega_{0}$ is at most one fourth of the angular frequency of the field $\omega_{1}$. Moreover, the relative error is less than $1\%$ as soon as $\epsilon \leq 0.033$. For comparison, the solution in the RWA is quite bad, since $E_{R}^{RWA}(\epsilon)$ has a relative error less than $15\%$ only when $\epsilon \leq 0.07$. Therefore, we conclude that both the two-term approximate solution given in (\ref{Aprox2terminoU}) and the two-term normalized approximate solution given in (\ref{Aprox2terminoU}) and (\ref{Paso2})-(\ref{Paso1b}) are accurate descriptions of the system from $t= 0$ to  $t = 10(2\pi/\Omega_{0})$, that is, during the first $10$ Rabi oscillations as long as $\Omega_{0}  \leq \omega_{1}/4$.

%%%%%%%%%%%%%%%%%%%%%%%%%%%%%%%%%%
%%%%%%%%%%%%%%%%%%%%%%%%%%%%%%%%%%
%%%%%%%%%%%%%%%%%%%%%%%%%%%%%%%%%%
%%%%%%%%%%%%%%%%%%%%%%%%%%%%%%%%%%
\subsection{Special cases}

In this section we present two classes of initial conditions that lead to a large difference between the two-term multiple-scales solution and the solution in the RWA.  

First assume that the qubit is initially in the ground state $\vert 1 \rangle$. It follows that $\rho_{I}(0) = \rho (0) = \vert 1 \rangle \langle 1 \vert$,  since the Schr\"{o}dinger picture and the IP coincide at $t=0$. Then, from (\ref{Matriz2}) and (\ref{Matriz3}) one has $\rho_{12}(0) = 0$ and $\alpha_{30}(0) = -1$. Substituting these values in (\ref{Aprox2terminoU}) one obtains that
\begin{eqnarray}
\label{Aprox2terminoUE}
\rho_{12}(t) &=& -  \frac{i}{2} \mbox{sin}\left( \Omega_{0}t \right) \cr
&& - \frac{\epsilon}{2} \left[ \mbox{cos}^{2}\left( \frac{\Omega_{0}t}{2} \right) - e^{-i2\omega_{1}t}\mbox{cos}(\Omega_{0}t)  \right]  \ , \cr
&& \cr
\alpha_{30}(t) &=& -\mbox{cos}\left(\Omega_{0}t\right) + \epsilon  \mbox{sin}(\Omega_{0}t)\mbox{sin}(2\omega_{1}t)  \ .
\end{eqnarray}
Hence, the other two components of the Bloch vector (\ref{ComponentesBloch}) are given by
\begin{eqnarray}
\label{BlochE}
\alpha_{10}(t) &=& \epsilon \left[ -\mbox{cos}^{2}\left( \frac{\Omega_{0}t}{2} \right) + \mbox{cos}(2\omega_{1}t)\mbox{cos}(\Omega_{0}t)  \right]  \ , \cr
&& \cr
\alpha_{20}(t) &=&  - \mbox{sin}\left( \Omega_{0}t \right) - \epsilon \mbox{sin}(2\omega_{1}t)\mbox{cos}(\Omega_{0}t)   \ , 
\end{eqnarray}
while the probability to find the qubit in the excited state is
\begin{eqnarray}
\label{probabilidad}
\rho_{22}(t) &=& \mbox{sin}^{2}\left( \frac{\Omega_{0}t}{2}\right) + \frac{\epsilon}{2}  \mbox{sin}(\Omega_{0}t)\mbox{sin}(2\omega_{1}t) \ .
\end{eqnarray}
First observe that the probability $\rho_{22}(t)$ to find the qubit in the excited state is modified from a sinusoidal oscillation described by sin$^{2}(\Omega_{0}t/2)$ to one that has modulations described by the second addend in (\ref{probabilidad}). This second term includes not only the Rabi frequency $\Omega_{0}$ but also the frequency of the field $\omega_{1}$. In addition, notice that this correction tends to zero as $\epsilon = \Omega_{0}/(2\omega_{1})$ tends to zero.

\begin{figure}
     \centering
     \subfloat[]{\includegraphics[scale=0.45]{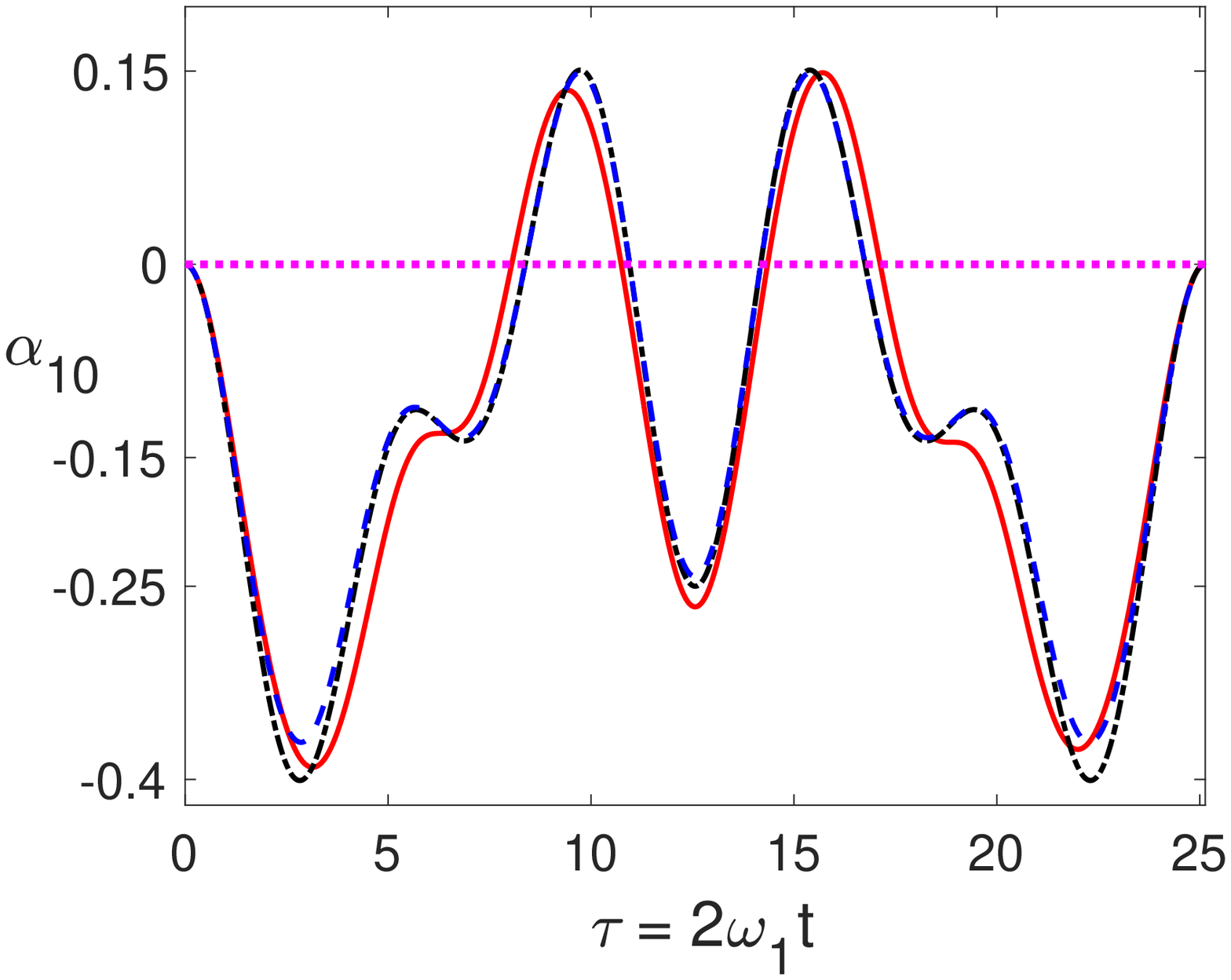}\label{Figure2a}}\\
     \subfloat[]{\includegraphics[scale=0.45]{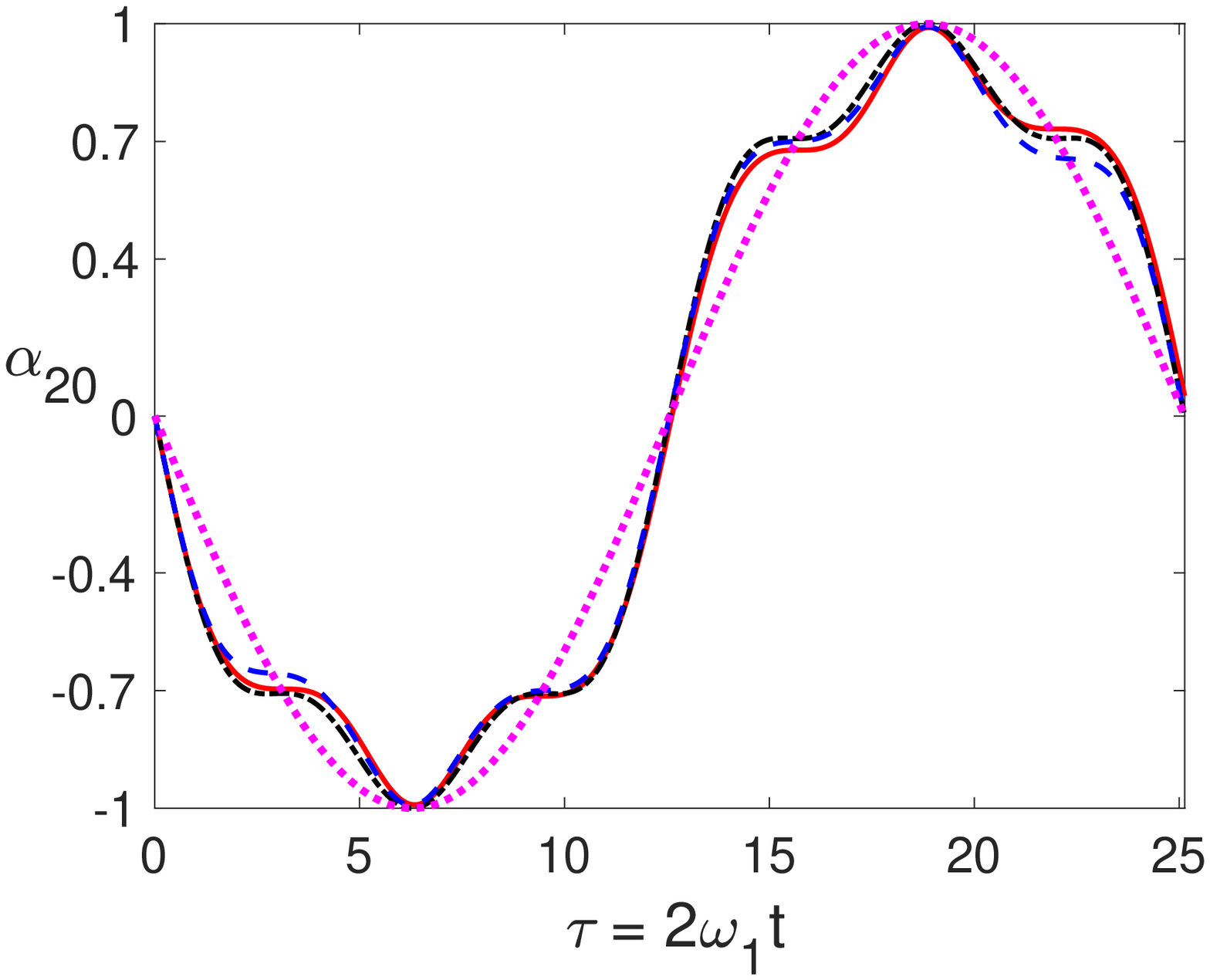}\label{Figure2b}}\\
     \subfloat[]{\includegraphics[scale=0.45]{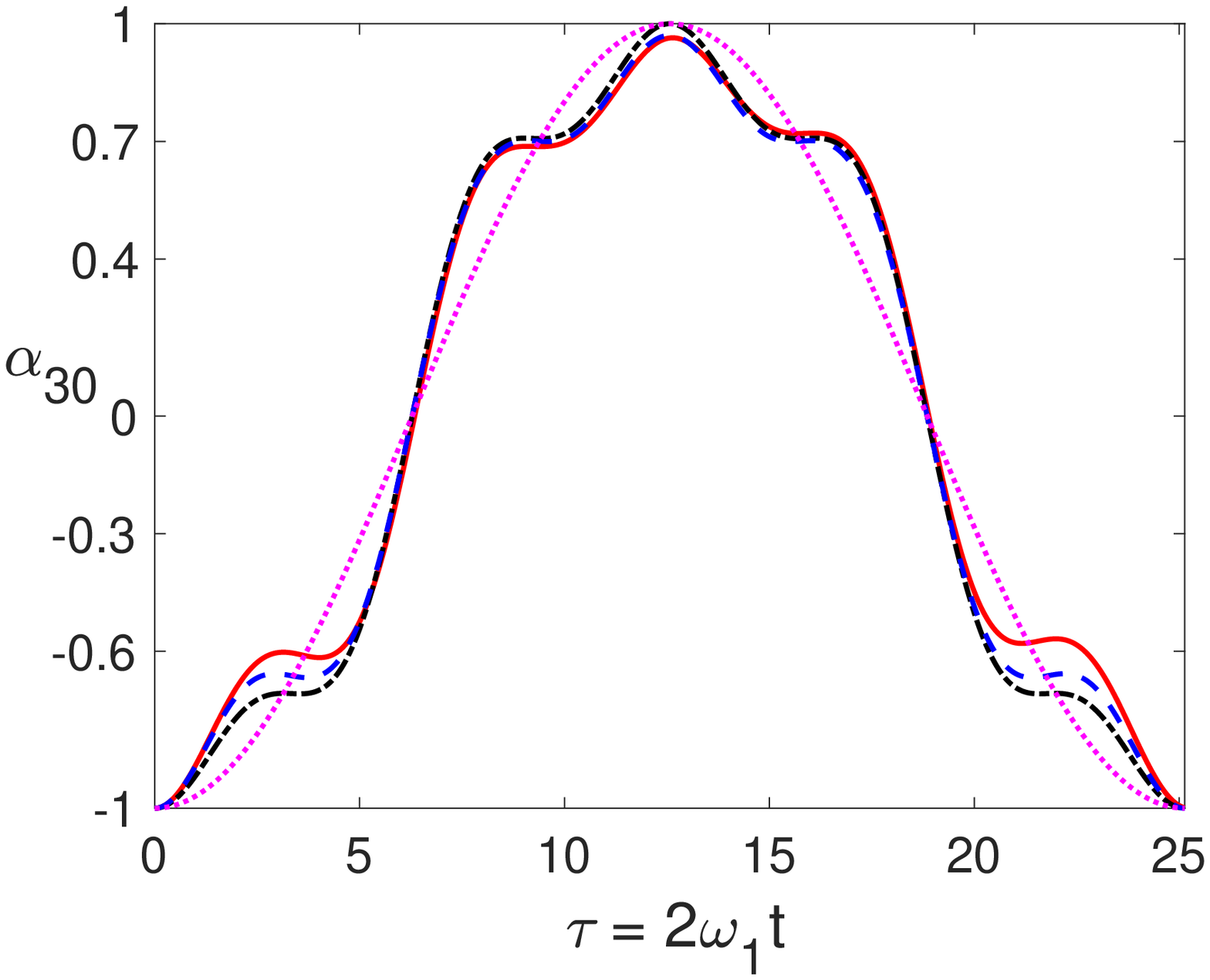}\label{Figure2c}}
     \caption{The components $\alpha_{10}$, $\alpha_{20}$, and $\alpha_{30}$ of the Bloch vector are shown as a function of the non dimensional time $\tau = 2\omega_{1}t$ from $\tau = 0$ to $\tau = 2\pi /\epsilon$ with $\epsilon = 1/4$. The system is initially in the ground state $\vert 1 \rangle$.  The numerical solution is shown in red-solid lines, the two-term approximation in black-dot-dashed lines, the two-term normalized approximation in blue-dashed lines, and the solution in the RWA in magenta-dotted lines. Figs. \ref{Figure2a}, \ref{Figure2b}, and \ref{Figure2c} show $\alpha_{10}$, $\alpha_{20}$, and $\alpha_{30}$, respectively.}
     \label{Figure2}
\end{figure}

Figure \ref{Figure2} shows the comparison between the exact Bloch vector (red-solid line) and the two-term approximate Bloch vector (black-dot-dashed line) in (\ref{Aprox2terminoUE}) and (\ref{BlochE}) as a function of the non-dimensional time $\tau = 2\omega_{1}t$. It also shows the comparison with the normalized two-term approximate Bloch vector (blue-dashed line) and the Bloch vector in the RWA (magenta-dotted line). The exact Bloch vector was obtained from (\ref{Matriz3})-(\ref{ComponentesBloch}) and the numerical solution of von Neumann's equation (\ref{13}) and (\ref{15}), while the normalized two-term approximate Bloch vector was obtained by normalizing the approximate Bloch vector in  (\ref{Aprox2terminoUE}) and (\ref{BlochE}) and the Bloch vector in the RWA was obtained from (\ref{TE36Bloch}).  The comparison is made for $\epsilon = 1/4$  and from $\tau = 0$ to  $\tau = 2\pi / \epsilon$, that is, during the first Rabi oscillation. Notice that the agreement between the numerical and the two-term approximate solution is remarkable, even though the Rabi frequency $\Omega_{0}$ is just half the angular frequency of the field $\omega_{1}$, see the definition of $\epsilon$ in  (\ref{eps0}). In addition, observe that the normalized approximate solution is even better and that the solution in the RWA is quite bad, especially for the value of $\alpha_{10}$. Finally, we note that the relative error between the exact Bloch vector and the (normalized or non normalized) two-term approximate Bloch vector is less than $25\%$ in the time interval $0 \leq \tau = 2\omega_{1}t \leq 3(2\pi/\epsilon)$, that is, during the first three Rabi oscillations.

In order to explain the differences we rely on the geometric description presented at the beginning of Sec. IV. The initial condition $\rho_{I}(0) = \rho (0) = \vert 1 \rangle \langle 1 \vert$ pertains to the class of initial conditions where the Bloch vector $\mathbf{r}_{I}(0)$ is located in the $yz$-plane:
\begin{eqnarray}
\label{CondicionInicialCaso1}
\mathbf{r}_{I}(0) &=& \left( \alpha_{10}(0), \alpha_{20}(0), \alpha_{30}(0) \right)^{T}  \ ,  \cr 
&=& \left( 0, \alpha_{20}(0), \alpha_{30}(0) \right)^{T} \   .
\end{eqnarray}
Substituting (\ref{CondicionInicialCaso1}) into (\ref{TE36Bloch}) and (\ref{TE43}) one obtains
\begin{eqnarray}
\label{RWAcaso1}
\alpha_{10}^{\mbox{\tiny RWA}}(t) &=&  0 \ , \cr 
&& \cr
\alpha_{20}^{\mbox{\tiny RWA}}(t) &=& \mbox{cos}(\Omega_{0}t)\alpha_{20}(0) + \mbox{sin}(\Omega_{0}t)\alpha_{30}(0) \ ,\cr
&& \cr
\alpha_{30}^{\mbox{\tiny RWA}}(t) &=& -\mbox{sin}(\Omega_{0}t)\alpha_{20}(0) + \mbox{cos}(\Omega_{0}t) \alpha_{30}(0) \ , \cr
&& 
\end{eqnarray}
and
\begin{eqnarray}
\label{DTcaso1}
\alpha_{10}(t) &=&  -\frac{\epsilon}{2}\mbox{sin}(\Omega_{0}t) \left[ 1 - 2\mbox{cos}(2\omega_{1}t) \right] \alpha_{20}(0) \cr
&& + \frac{\epsilon}{2}\left\{ 1 + \mbox{cos}(\Omega_{0}t) \left[ 1 - 2\mbox{cos}(2\omega_{1}t) \right]  \right\} \alpha_{30}(0) \ , \cr
&& \cr
\alpha_{20}(t) &=& \alpha_{20}^{\mbox{\tiny RWA}}(t) \cr
&&  -\epsilon \mbox{sin}(\Omega_{0}t) \mbox{sin}(2\omega_{1}t) \alpha_{20}(0) \cr
&& +\epsilon \mbox{cos}(\Omega_{0}t) \mbox{sin}(2\omega_{1}t)\alpha_{30}(0) \ , \cr
&& \cr
\alpha_{30}(t) &=& \alpha_{30}^{\mbox{\tiny RWA}}(t) \cr
&& -\epsilon \mbox{sin}(2\omega_{1}t) \mbox{cos}(\Omega_{0}t) \alpha_{20}(0) \cr
&& -\epsilon \mbox{sin}(2\omega_{1}t) \mbox{sin}(\Omega_{0}t) \alpha_{30}(0) \ .
\end{eqnarray}
One immediately finds that the rotating wave approximation predicts that $\alpha_{10}^{\mbox{\tiny RWA}}(t)  =0 $, while the two-term approximation establishes that this is not true. To explain the differences between the two solutions recall that in the RWA the Bloch vector simply rotates around the $x$-axis with angular velocity $-\Omega_{0}$ (recall that a negative angular velocity indicates a clockwise rotation). Since the initial condition in (\ref{CondicionInicialCaso1}) describes a Bloch vector initially contained in the $yz$-plane, it follows that in the RWA the Bloch vector will always be contained in the $yz$-plane because it only rotates around the $x$-axis. On the other hand, the two-term multiple-scales solution in (\ref{DTcaso1}) not only describes the precessional motion of the Bloch vector, but the also the nutational one. Therefore, the two-term multiple-scales solution in (\ref{DTcaso1}) indicates that the Bloch vector is, in general, not contained in the $yz$-plane due to the nutational motion. Moreover, this difference between the RWA and real solution holds for all times, since it was established at the beginning of Sec. IV that the true evolution of the Bloch vector describes a precessional-nutational motion. Also, the two-term multiple-scales solution is able to describe accurately this true evolution for long times according to the results in Sec. IVD, see Fig. \ref{Figure1}.

\begin{figure}
     \centering
     \subfloat[]{\includegraphics[scale=0.45]{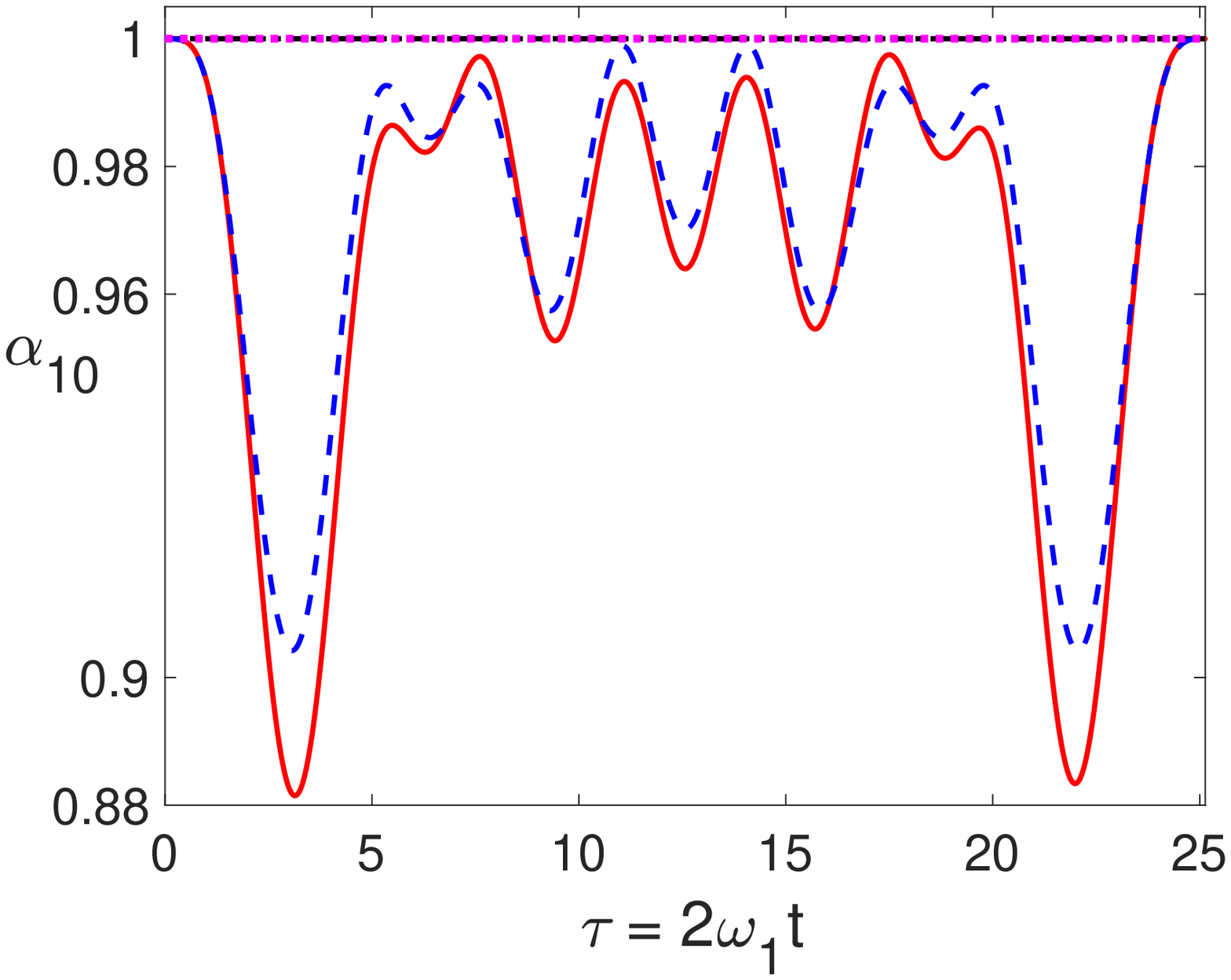}\label{Figure3a}}\\
     \subfloat[]{\includegraphics[scale=0.45]{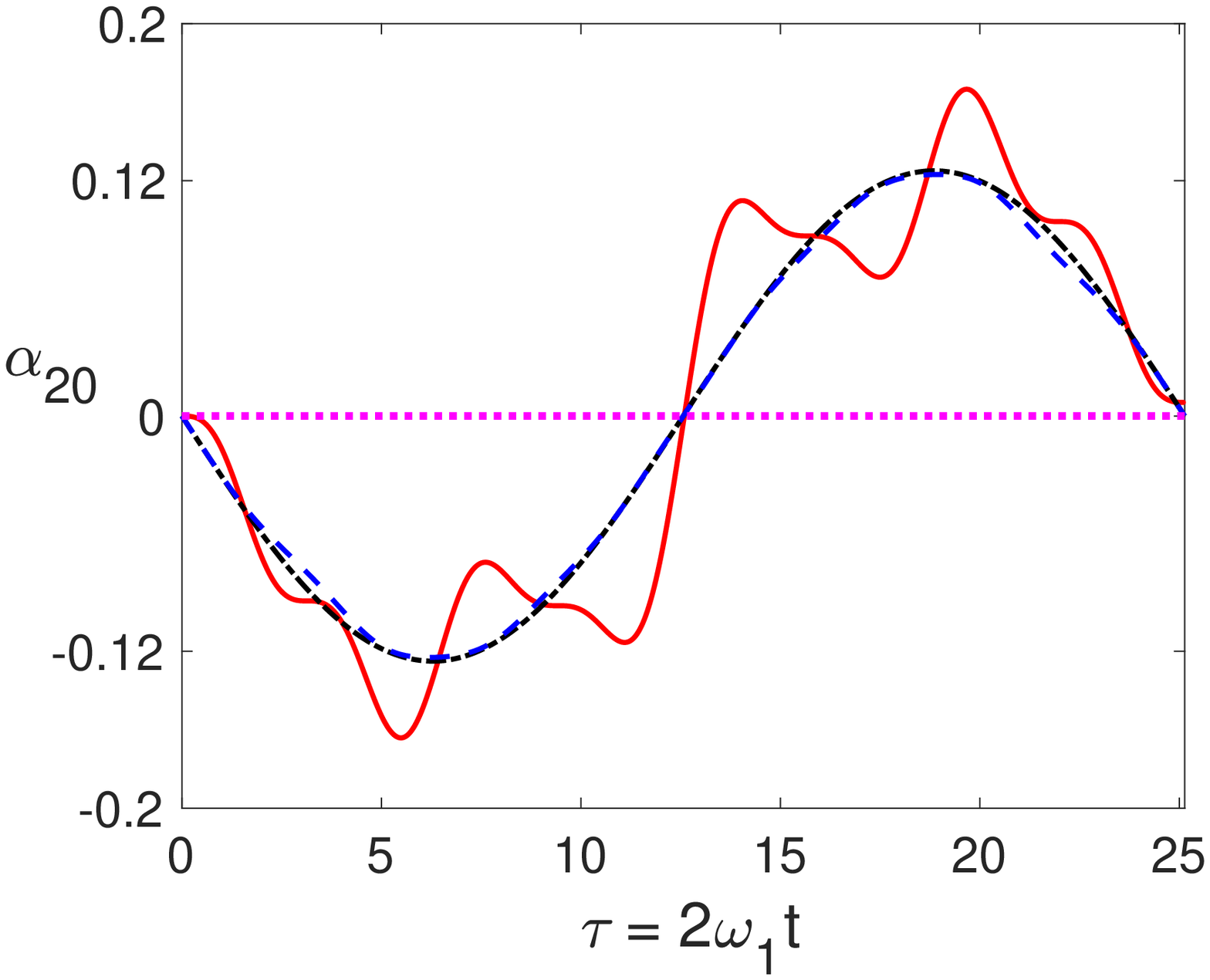}\label{Figure3b}}\\
     \subfloat[]{\includegraphics[scale=0.45]{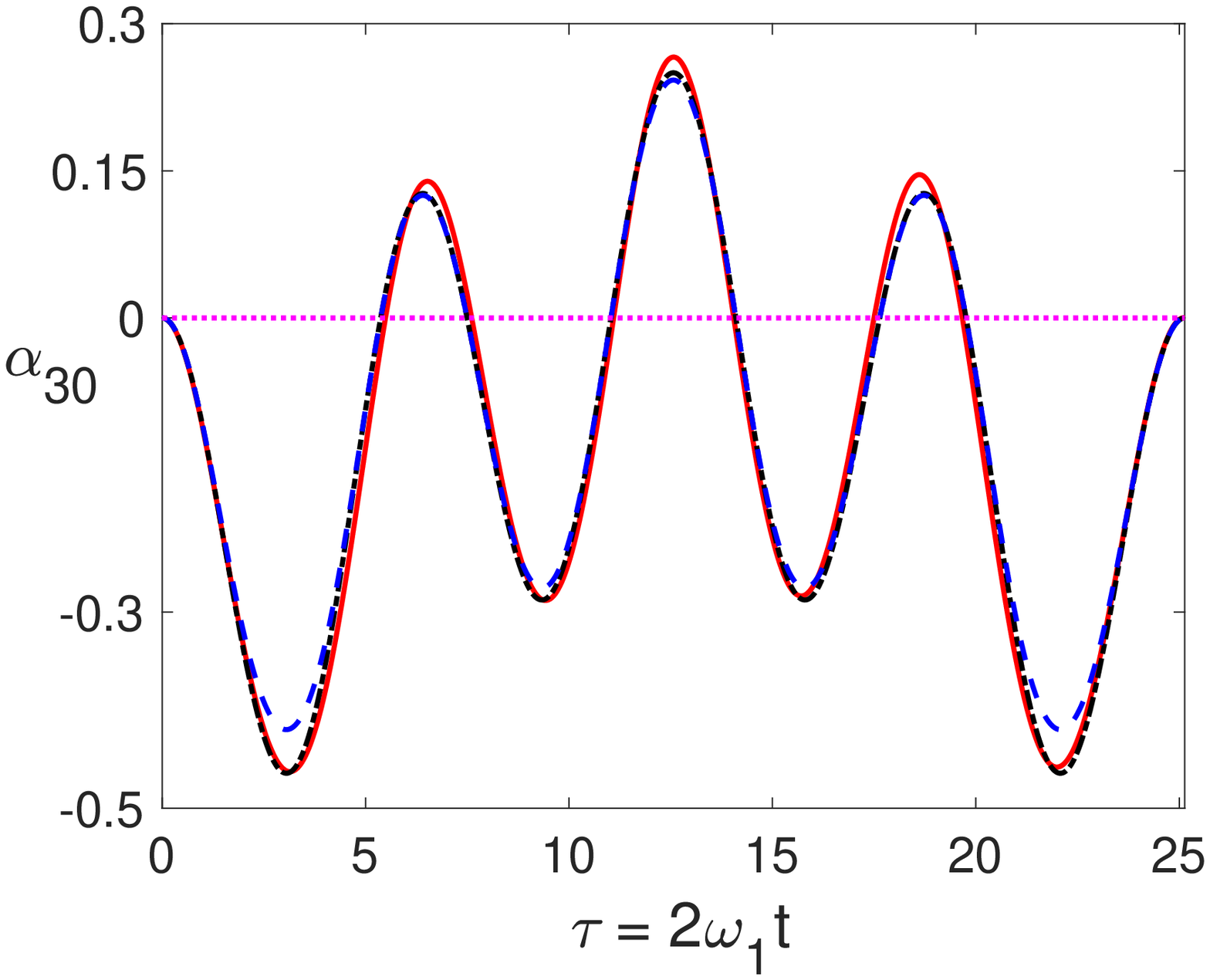}\label{Figure3c}}
     \caption{The components $\alpha_{10}$, $\alpha_{20}$, and $\alpha_{30}$ of the Bloch vector are shown as a function of the non dimensional time $\tau = 2\omega_{1}t$ from $\tau = 0$ to $\tau = 2\pi /\epsilon$ with $\epsilon = 1/4$. The system is initially in the state $\vert \phi \rangle = \left( \vert 1 \rangle + \vert 2 \rangle \right)/\sqrt{2}$.  The numerical solution is shown in red-solid lines, the two-term approximation in black-dot-dashed lines, the two-term normalized approximation in blue-dashed lines, and the solution in the RWA in magenta-dotted lines. Figs. \ref{Figure3a}, \ref{Figure3b}, and \ref{Figure3c} show $\alpha_{10}$, $\alpha_{20}$, and $\alpha_{30}$, respectively.}
     \label{Figure3}
\end{figure}

Now assume that the qubit is initially in the (normalized) state  $\vert \phi \rangle = (1/\sqrt{2})\left( \vert 1 \rangle + \vert 2 \rangle \right)$, which corresponds to a linear superposition of the excited and ground states with equal coefficients. It follows that the density operator in the IP has the matrix representation with respect to the basis $\gamma = \{ \vert 1 \rangle, \ \vert 2 \rangle \}$ given by $[ \rho_{I}(0) ]_{\gamma} = [ \rho (0) ]_{\gamma} = (1/2)\mathbb{I}_{2}$ with $\mathbb{I}_{2}$ the $2 \times 2$ identity matrix,  since the Schr\"{o}dinger picture and the IP coincide at $t=0$. Then, from (\ref{Matriz2}) and (\ref{Matriz3}) one has $\rho_{12}(0) = 1/2$ and $\alpha_{30}(0) = 0$. Using the definition of the Bloch vector in (\ref{Bloch}) and (\ref{ComponentesBloch}) it follows that $\alpha_{10}(0) = 1$ and $\alpha_{20}(0) = 0$. Substituting these values in (\ref{TE36Bloch}) and (\ref{TE43}) one obtains 
\begin{eqnarray}
\label{RWAcaso2}
\mathbf{r}_{I}^{\mbox{\tiny RWA}}(t) &=&  (1 , 0 , 0)^{T} \ ,
\end{eqnarray}
and the two-term multiple-scales solution
\begin{eqnarray}
\label{DTcaso2}
\alpha_{10}(t) &=& 1  \ , \cr
&& \cr
\alpha_{20}(t) &=&  -\frac{\epsilon}{2}\mbox{sin}(\Omega_{0}t)  \ , \cr
&& \cr
\alpha_{30}(t) &=& -\frac{\epsilon}{2}\left[ 1 - 2\mbox{cos}(2\omega_{1}t) + \mbox{cos}(\Omega_{0}t) \right]  \  .
\end{eqnarray}
Comparing (\ref{RWAcaso2}) and (\ref{DTcaso2}) one immediately recognizes that the Bloch vector in the RWA is very different from the Bloch vector in the two-term approximation. Explicitly, Figure \ref{Figure3} shows the comparison between the exact Bloch vector (red-solid line) and the two-term approximate Bloch vector (black-dot-dashed line) in (\ref{DTcaso2}) as a function of the non-dimensional time $\tau = 2\omega_{1}t$. It also shows the comparison with the normalized two-term approximate Bloch vector (blue-dashed line) and the Bloch vector in the RWA (magenta-dotted line). The exact Bloch vector was obtained from (\ref{Matriz3})-(\ref{ComponentesBloch}) and the numerical solution of von Neumann's equation (\ref{13}) and (\ref{15}), while the normalized two-term approximate Bloch vector was obtained by normalizing the approximate Bloch vector in  (\ref{DTcaso2}) and the Bloch vector in the RWA was obtained from (\ref{RWAcaso2}).  The comparison is made for $\epsilon = 1/4$  and from $\tau = 0$ to  $\tau = 2\pi / \epsilon$, that is, during the first Rabi oscillation. Notice that the agreement between the numerical and the normalized two-term approximate solution is quite good, even though the Rabi frequency $\Omega_{0}$ is just half the angular frequency of the field $\omega_{1}$, see the definition of $\epsilon$ in  (\ref{eps0}). In addition, observe that the solution in the RWA is very bad, while the two-term approximation only describes accurately $\alpha_{30}(t)$. Finally, we note that the relative error between the exact Bloch vector and the (normalized or non normalized) two-term approximate Bloch vector is less than $15\%$ in the time interval $0 \leq \tau = 2\omega_{1}t \leq 10(2\pi/\epsilon)$, that is, during the first ten Rabi oscillations.

The initial condition $\vert \phi \rangle = (1/\sqrt{2})\left( \vert 1 \rangle + \vert 2 \rangle \right)$ used above pertains to the class of initial conditions where the Bloch vector $\mathbf{r}_{I}(0)$ is located along the $x$-axis:
\begin{eqnarray}
\label{CondicionInicialCaso2}
\mathbf{r}_{I}(0) &=& \left( \alpha_{10}(0), 0, 0 \right)^{T}  \   .
\end{eqnarray}
Substituting (\ref{CondicionInicialCaso2}) into (\ref{TE36Bloch}) and (\ref{TE43}) one obtains
\begin{eqnarray}
\label{RWAcaso2G}
\mathbf{r}_{I}^{\mbox{\tiny RWA}}(t) &=&  ( \alpha_{10}(0)  , 0 , 0)^{T} \  ,
\end{eqnarray}
and the two-term multiple-scales solution
\begin{eqnarray}
\label{DTcaso2G}
\alpha_{10}(t) &=& \alpha_{10}(0)  \ , \cr
&& \cr
\alpha_{20}(t) &=&  -\frac{\epsilon}{2}\mbox{sin}(\Omega_{0}t)\alpha_{10}(0)  \ , \cr
&& \cr
\alpha_{30}(t) &=& -\frac{\epsilon}{2}\left[ 1 - 2\mbox{cos}(2\omega_{1}t) + \mbox{cos}(\Omega_{0}t) \right] \alpha_{10}(0) \  .
\end{eqnarray}
Why is the solution in the RWA so different? Well, since the initial condition in (\ref{CondicionInicialCaso2}) describes a Bloch vector that initially points along the $x$-axis, it follows that in the RWA the Bloch vector will always be contained in the $x$-axis because it only rotates around the $x$-axis. On the other hand, the two-term multiple-scales solution in (\ref{DTcaso2G}) indicates that the Bloch vector is, in general, not contained in the $x$-axis due to the nutational motion. Again, this difference between the RWA and the true solution holds for all times, since it was established at the beginning of Sec. IV that the true evolution of the Bloch vector describes a precessional-nutational motion. In particular, the two-term multiple-scales solution describes accurately this true evolution for long times according to the results in Sec. IVD, see Fig. \ref{Figure1}.

\section{Conclusions}

In this article we considered the semiclassical Rabi model with the condition of resonance, that is, a two-level system (a qubit) interacting with a classical, single-mode field with angular frequency $\omega_{1}$ such that the field is resonant with the qubit's transition.  The time evolution of this system is usually solved with the rotating-wave-approximation (RWA) where the counterrotating terms are neglected. It leads to a simple analytic solution that is easy to interpret physically and that provides accurate results when the Rabi frequency (or qubit-field coupling) $\vert \Omega_{0} \vert$ is much smaller than the frequency of the field. An alternative to the RWA is to use the method of multiple-scales. This method allows one to include the counterrotating terms neglected in the RWA and also leads to simple, approximate, and analytic solutions that are much more accurate for very long times and that provide physical insight to the evolution of the system. In particular, we found that the multiple-scales solution accurately describes the evolution of the Bloch vector of the system like a \textit{nutational-precessional motion}  similar to the motion of a symmetric top with one point fixed. In great measure, the rotating terms are responsible for the precessional motion, while the counterrotating terms are responsible for the nutational motion.

We determined that, for any pure-state initial condition, the relative error between the exact and the two-term multiple scales solution presented in the article is less than $15\%$ if $\omega_{1} \geq 2\vert \Omega_{0} \vert$ and one Rabi oscillation is considered. One also has a relative error less than $15\%$ if $\omega_{1} \geq 4\vert \Omega_{0} \vert$ and 10 Rabi oscillations are considered. These results indicate that the two-term multiple-scales solution presented in this article describes to good approximation the system even when the angular frequency of the field is not much larger than the Rabi frequency.

In addition, the approximate, analytic solutions indicate the corrections to the solution in the RWA and provide a quantitative criterion to determine when the solution in the RWA is accurate. The corrections consist of terms that oscillate at the angular frequency of the field $\omega_{1}$ and the Rabi frequency $\vert \Omega_{0} \vert$, while the criterion indicates that the absolute value of the corrections to the solution in the RWA are $\leq  2\vert \Omega_{0} \vert /\omega_{1}$. In terms of the Bloch vector, the corrections introduce the nutational motion.

Finally, the multiple scales method can be used to include the counterrotating terms in a system composed of many two-level systems and in a system where there is a quantum field instead of a classical one (the Jaynes-Cummings and Tavis-Cummings models). This is work in progress.

%%%%%%%%%%%
%%%%%%%%%%%
%%%%%%%%%%%
%%%%%%%%%%%
\appendix

\section{Two time-scales and the RWA}

In this appendix we solve von Neumann's equation (\ref{13}) with the interaction Hamiltonian in (\ref{15}) in two forms. First, using the rotating-wave-approximation (RWA), and then using the multiple-scales method with the two time-scales in (\ref{2E}). Before this is done it is appropriate to express everything in terms of non-dimensional quantities.

\subsection{Non-dimensional quantities}

In the rest of this appendix we measure time in units of $1/(2\omega_{1})$ and we introduce the following quantities:
\begin{eqnarray}
\label{16}
\tau &=& 2\omega_{1}t \ , \quad\quad\quad\quad \ \epsilon = \frac{\Omega_{0}}{2\omega_{1}} \ , \cr
&& \cr
\tilde{\rho}_{I}(\tau) &=& \rho_{I}\left( \frac{\tau}{2\omega_{1} } \right) \ , \quad\quad \alpha_{30}(\tau)  \ = \ \rho_{22}(\tau) - \rho_{11}(\tau) \ , \cr
&& \cr
\rho_{\lambda \lambda '}(\tau) &=& \langle \lambda \vert \tilde{\rho}_{I}(\tau) \vert \lambda ' \rangle \ , \quad (\lambda, \lambda' = 1,2)  \ .
\end{eqnarray}
Notice that $\tau$ is the \textit{non-dimensional time} and that $\epsilon$ is exactly the same as in (\ref{eps0}). Also, $\tilde{\rho}_{I}(\tau)$ is the density operator of the system in the IP as a function of the non-dimensional time $\tau$, $ \rho_{\lambda \lambda '}(\tau)$ is a matrix element of  $\tilde{\rho}_{I}(\tau)$ in the basis $\gamma$ defined in (\ref{1Base}), and $\alpha_{30}(\tau)$ is the probability of finding the qubit in the excited state minus the probability of finding the qubit in the ground state. Finally, the values of $\rho_{\lambda \lambda '}(\tau)$ and $\alpha_{30}(\tau)$ are exactly the same as those of $\rho_{\lambda \lambda'}(t)$ and $\alpha_{30}(t)$ in (\ref{Matriz2}) and (\ref{Matriz3}). The difference is that the latter are considered as functions of time $t$, while the former are considered as functions of the non dimensional time $\tau$. We have used the same symbol for the quantities and they are distinguished by the variable $t$ or $\tau$. Also, $\rho_{\lambda \lambda '}(\tau)$ and $\alpha_{30}(\tau)$ only appear in the appendix  and  $\rho_{\lambda \lambda'}(t)$ and $\alpha_{30}(t)$ only appear in the main text.

Before proceeding we introduce matrices that are going to be used throughout the next sections. We prefer to group them together so that the reader can make easy reference to all of them. We define the matrices
\begin{eqnarray}
\label{DefinicionMatrices}
\mathbf{X}(\tau) &=& \left( \rho_{12}(\tau), \rho_{21}(\tau) , \alpha_{30}(\tau) \right)^{T} \ , \cr 
&& \cr
\mathbb{A}_{0} &=& \frac{i}{2} \left(
\begin{array}{ccc}
0 & 0                   & 1 \cr
0                   & 0  & -1 \cr
2  & -2  & 0
\end{array}
\right) \ , \cr
&& \cr
&& \cr
\mathbb{A}_{1}(t_{1}) &=& \frac{i}{2} \left(
\begin{array}{ccc}
0                   & 0                   & e^{-it_{1}} \cr
0                   & 0                   & -e^{it_{1}} \cr
2e^{it_{1}}  & -2e^{-it_{1}}  & 0
\end{array}
\right) \ , \cr
&& \cr
&& \cr
\mathbb{A}_{2}(t_{1}) &=& \frac{1}{2}\left(
\begin{array}{ccc}
0 & 0 & -e^{-it_{1}} \cr
0 & 0 & -e^{it_{1}} \cr
2e^{it_{1}} & 2e^{-it_{1}} & 0 
\end{array}
\right) \ , \cr
&& \cr
&& \cr
\mathbb{A}_{3} &=& \frac{i}{2}\left(
\begin{array}{ccc}
1 & 0 & 0 \cr
0 & -1 & 0 \cr
0 & 0 & 0
\end{array}
\right) \ , \cr
&& \cr
&& \cr
\mathbb{A}_{4}(t_{1}) &=& -\frac{1}{4} \left(
\begin{array}{ccc}
0 & e^{-i2t_{1}} & 0 \cr
e^{i2t_{1}}& 0 & 0 \cr
0 & 0 & 0
\end{array}
\right) \ , \cr
&& \cr
&& \cr
\mathbb{Q} &=& \left(
\begin{array}{ccc}
1 & -1 & 1 \cr
-1 & 1 & 1 \cr
2 & 2 & 0 
\end{array}
\right) \ , \cr
&& \cr
&& \cr
\mathbb{D} &=& \left(
\begin{array}{ccc}
i & 0  & 0 \cr
0 & -i & 0 \cr
0 & 0 & 0  
\end{array}
\right) \ , \cr
&& \cr
&& \cr
e^{\mathbb{D}t_{2}} &=& \left(
\begin{array}{ccc}
e^{it_{2}} & 0  & 0 \cr
0 & e^{-it_{2}} & 0 \cr
0 & 0 & 1  
\end{array}
\right) \ , \cr
&& \cr
&& \cr
\mathbb{B} &=& \mathbb{Q}^{-1} \Big( \mathbb{A}_{3} -  [ \mathbb{A}_{0} , \mathbb{A}_{2}(0) ] \Big) \mathbb{Q} \  , \cr
&& \cr
&& \cr
\mathbb{G}_{1}(t_{2}) &=& \frac{1}{2}\left( 
\begin{array}{ccc}
-i\mbox{sin}(t_{2}) & 0 & \mbox{sin}^{2}(t_{2}/2) \cr
0 & i\mbox{sin}(t_{2}) & \mbox{sin}^{2}(t_{2}/2) \cr
2\mbox{sin}^{2}(t_{2}/2) & 2\mbox{sin}^{2}(t_{2}/2) & 0 
\end{array}
\right) \ , \cr
&& \cr
&& \cr
\mathbb{W}(t_{1},t_{2}) &=& \mathbb{G}_{1}(t_{2}) e^{-\mathbb{A}_{0}t_{2}} + \mathbb{A}_{2}(t_{1}) - \mathbb{A}_{2}(0) \   , \cr
&& \cr
&& \cr
\mathbb{U} &=& \left( 
\begin{array}{ccc}
1 & 1 & 0 \cr
-i & i & 0 \cr
0 & 0 & 1
\end{array}
\right) \ .
\end{eqnarray}
Here $T$ indicates the transpose, $[ \cdot , \cdot ]$ is the commutator,  and $\tau$, $t_{1}$, and $t_{2}$ are arbitrary real numbers. Also, $\mathbb{Q}$ is a matrix whose columns are eigenvectors of $\mathbb{A}_{0}$ and $\mathbb{D}$ is a diagonal matrix with the eigenvalues of $\mathbb{A}_{0}$ on the diagonal and in the order given by the columns of $\mathbb{Q}$ so that
\begin{eqnarray}
\label{factorizacion}
\mathbb{A}_{0} = \mathbb{Q}\mathbb{D} \mathbb{Q}^{-1} \ , \quad e^{\mathbb{A}_{0}t_{2}} = \mathbb{Q}e^{\mathbb{D}t_{2}} \mathbb{Q}^{-1} \ .
\end{eqnarray}

Using (\ref{16}) one can write von Newmann's equation in the IP given in (\ref{13})  as a system of linear, first order differential equations:
\begin{eqnarray}
\label{25}
\frac{d}{d\tau} \mathbf{X}(\tau) &=& \epsilon \left[ \mathbb{A}_{0} + \mathbb{A}_{1}(\tau) \right] \mathbf{X}(\tau) \ , 
\end{eqnarray}
with $\mathbf{X}(\tau)$, $\mathbb{A}_{0}$, and $\mathbb{A}_{1}(\tau)$ in (\ref{DefinicionMatrices}). 

Now, the first step is to solve (\ref{25}) in the rotating-wave-approximation (RWA), since this provides the necessary motivation for the multiple-scales solution presented in the following sections.

\subsection{The rotating-wave-approximation (RWA)}

In the RWA one assumes that $\epsilon \ll 1$ and that $\tilde{\rho}_{I}(\tau)$ evolves appreciably in a time-scale much larger than $1$. Then, the terms in (\ref{25}) that are multiplied by $\mathbb{A}_{1}(\tau)$ average to zero because they have a time dependence of the form $e^{\pm i \tau}$ and evolve appreciably in a time-scale of $1$. Hence, one can neglect the terms multiplied by $\mathbb{A}_{1}(\tau)$ and (\ref{25}) reduces to
\begin{eqnarray}
\label{RWAeq}
\frac{d}{d\tau} \mathbf{X}(\tau) &=& \epsilon \mathbb{A}_{0} \mathbf{X}(\tau) \ .
\end{eqnarray}
This is a system of linear, first order equations with constant coefficients and, thus, can be solved exactly by calculating the eigenvectors and eigenvalues of $\mathbb{A}_{0}$. The general solution of (\ref{RWAeq}) is 
\begin{eqnarray}
\label{RWA}
\mathbf{X}(\tau ) &=& e^{\epsilon \mathbb{A}_{0} \tau} \mathbf{X}(0) = \mathbb{Q} e^{\epsilon \mathbb{D} \tau}\mathbb{Q}^{-1} \mathbf{X}(0) \ ,
\end{eqnarray}
where $\mathbb{A}_{0}$, $\mathbb{Q}$, and $e^{\epsilon \mathbb{D} \tau}$ are given in (\ref{DefinicionMatrices}). Notice that we used (\ref{factorizacion}) in the second equality in (\ref{RWA}). Also, from (\ref{DefinicionMatrices}) and (\ref{RWA})  observe that the time-dependence of $\mathbf{X}(\tau)$ is of the form $e^{\pm i\epsilon\tau}$, so that the components of $\mathbf{X}(\tau)$ are periodic functions with period $2\pi/\epsilon$, which is the non-dimensional time for one Rabi oscillation. Carrying out the multiplication of matrices in (\ref{RWA}) and introducing units according to (\ref{16}) one obtains (\ref{RWAsolucionE}) in the main text.

To perform the RWA it was assumed that $\epsilon \ll 1$ and that $\tilde{\rho}_{I}(\tau)$ evolves appreciably in a time-scale much larger than $1$.  Since the time dependence of $\mathbf{X}(\tau)$ in (\ref{RWA}) is of the form $e^{\pm i\epsilon \tau}$, it follows that $\mathbf{X}(\tau)$  evolves on a time-scale of  $1/\epsilon$. Then, $\epsilon \ll 1$ guarantees that $\tilde{\rho}_{I}(\tau)$ evolves appreciably in a time-scale much larger than $1$. Hence, it is sufficient to ask that $\epsilon \ll 1$ for the RWA to hold.

Before ending this section it is important to note that the RWA can be performed directly in von Neumann's equation in (\ref{13}), as discussed in Section III. The relationship between both treatments is now briefly explained. Recall that the RWA corresponds to neglecting in (\ref{25}) the terms multiplied by $\mathbb{A}_{1}(\tau)$ to obtain (\ref{RWAeq}). In von Neumann's equation this corresponds to neglecting the counterrotating terms in $H_{II}^{0}(t)$ so that one is led to equation (\ref{15b}).

\subsection{The solution with two time-scales}

In this section we present the details of the multiple-scales method described in Sec. IVA. We express all quantities in matrix form to perform the calculations and we indicate the connection with the equations presented in Sec. IVA.

In the previous section it was found that the solution (\ref{RWA}) in the RWA holds as long as $\epsilon \ll 1$, that is, if $\epsilon$ is a perturbation parameter. In all that follows we assume that $\epsilon < 1$. From the discussion in the previous section and the geometric description presented in Sec. IVA, two time-scales that allow one to describe the evolution of $\mathbf{X}(\tau)$ induced by both $\mathbb{A}_{0}$ and $\mathbb{A}_{1}(\tau)$ are the following: 
\begin{eqnarray}
\label{40}
t_{1} = \tau \ , \quad\quad t_{2} = \epsilon \tau .
\end{eqnarray}
Here $t_{1}$ plays the role of the fast time-scale, while $t_{2}$ is the slow time-scale. Observe that the definitions of $t_{1}$ and $t_{2}$ in (\ref{2E}) are identical to those in (\ref{40}). Using the definition of $\tau$ in (\ref{16}) and the theory of multiple-scales \cite{Holmes}, the approximate solutions obtained using these two time-scales hold at least for times $\tau$ such that
\begin{eqnarray}
\label{Intervalo2}
0 \leq t_{2} = \epsilon \tau \leq \mathcal{O}(1) \quad \Leftrightarrow \quad 0 \leq t \leq \mathcal{O}\left( \frac{1}{\Omega_{0}} \right) \ .
\end{eqnarray}
This interval is exactly the same as that in (\ref{Intervalo2E}). Here and in the following $\mathcal{O}$ is the \textit{Big Oh} \cite{Holmes}. From the discussion in the previous section recall that $2\pi/\epsilon$ is the non dimensional time for one Rabi oscillation, so (\ref{Intervalo2}) indicates that the multiple-scales solution holds for $k>0$ Rabi oscillations. We now determine to good approximation the solution of (\ref{25}) using the two time-scales.

First define the function $\mathbf{Y}(t_{1},t_{2})$ by
\begin{eqnarray}
\label{41}
\mathbf{Y}\left[ t_{1}(\tau), t_{2}(\tau) \right] &=& \mathbf{X}(\tau) \  .
\end{eqnarray}
The connection between $\mathbf{Y}(t_{1},t_{2})$ given above and $\mathbf{K}(t_{1},t_{2})$ defined in (\ref{NewZ}) is the following:
\begin{eqnarray}
\label{ConexionYZ}
\mathbf{K}(t_{1}, t_{2}) &=& \mathbb{U} \mathbf{Y}(t_{1},t_{2}) \ ,
\end{eqnarray}
where the invertible matrix $\mathbb{U}$ is defined in (\ref{DefinicionMatrices}).

Substituting (\ref{41}) in (\ref{25}) one arrives at the initial value problem
\begin{eqnarray}
\label{43}
\frac{\partial \mathbf{Y}}{\partial t_{1}} (t_{1},t_{2}) &=& \epsilon \left[ \mathbb{A}_{0} + \mathbb{A}_{1}(t_{1}) \right]\mathbf{Y}(t_{1},t_{2}) - \epsilon \frac{\partial \mathbf{Y}}{\partial t_{2}}(t_{1},t_{2}) \ , \cr
&& \cr
\mathbf{Y}(0,0) &=& \mathbf{X}(0) \ ,
\end{eqnarray}
where we assume that $\mathbf{X}(0)$ does not depend on $\epsilon$. We note that the differential equation in (\ref{43}) is equivalent to the differential equation for $\mathbf{K}(t_{1},t_{2})$ in (\ref{TE6}) using (\ref{ConexionYZ}).

Now, assume that $\mathbf{Y}(t_{1},t_{2})$ has an asymptotic expansion of the form
\begin{eqnarray}
\label{44}
\mathbf{Y}(t_{1},t_{2}) \sim \mathbf{Y}_{0}(t_{1},t_{2}) + \epsilon \mathbf{Y}_{1}(t_{1},t_{2}) + \epsilon^{2}\mathbf{Y}_{2}(t_{1},t_{2}) + ... \ . \cr
&&
\end{eqnarray}
If one keeps $n$ terms in the righthand side of (\ref{44}), then one speaks of an $n$-term approximation for $\mathbf{Y}(t_{1},t_{2})$ and $\mathbf{X}(\tau)$. In what follows we are interested in obtaining a one- and two-term approximation of $\mathbf{X}(\tau)$. We note that the asymptotic expansion in (\ref{44}) is equivalent to the asymptotic expansion for $\mathbf{K}(t_{1},t_{2})$ in (\ref{TE7}) using (\ref{ConexionYZ}). Explicitly,
\begin{eqnarray}
\label{Conexion}
\mathbf{K}_{j}(t_{1},t_{2}) &=& \mathbb{U} \mathbf{Y}_{j}(t_{1},t_{2}) \ , \quad (j = 0,1, 2, ...).
\end{eqnarray}

Substituting (\ref{44}) in (\ref{43}) and equating equal powers of $\epsilon$, one arrives at the following first three initial value problems:
\begin{eqnarray}
\label{45}
&\mathcal{O}(1):& \quad \frac{\partial \mathbf{Y}_{0}}{\partial t_{1}}(t_{1},t_{2}) = \mathbf{0},  \cr
&&\cr
&& \quad \mathbf{Y}_{0}(0,0) = \mathbf{X}(0) \ . \cr
&& \cr
&\mathcal{O}(\epsilon):& \quad \frac{\partial \mathbf{Y}_{1}}{\partial t_{1}}(t_{1},t_{2}) = \left[  \mathbb{A}_{0} + \mathbb{A}_{1}(t_{1}) \right]\mathbf{Y}_{0}(t_{1},t_{2}) \cr
&& \quad\quad\quad\quad\quad\quad\quad\quad - \frac{\partial \mathbf{Y}_{0}}{\partial t_{2}}(t_{1},t_{2}) ,  \cr
&& \cr
&& \quad \mathbf{Y}_{1}(0,0) = \mathbf{0} \ . \cr
&& \cr
&\mathcal{O}(\epsilon^{2}):& \quad \frac{\partial \mathbf{Y}_{2}}{\partial t_{1}}(t_{1},t_{2}) = \left[  \mathbb{A}_{0} + \mathbb{A}_{1}(t_{1}) \right]\mathbf{Y}_{1}(t_{1},t_{2}) \cr
&& \quad\quad\quad\quad\quad\quad\quad\quad - \frac{\partial \mathbf{Y}_{1}}{\partial t_{2}}(t_{1},t_{2}) , \cr
&& \cr
&& \quad \mathbf{Y}_{2}(0,0) = \mathbf{0} \ . 
\end{eqnarray}
We note that the problems in (\ref{45}) are equivalent to those for $\mathbf{K}_{j}(t_{1},t_{2})$ in (\ref{TE8}) using (\ref{Conexion}).

We now solve the $\mathcal{O}(1)$ problem in (\ref{45}).  One immediately finds that
\begin{eqnarray}
\label{46}
\mathbf{Y}_{0}(t_{1},t_{2}) &=& \mathbf{Y}_{0}(0,t_{2}) \ .
\end{eqnarray}
Notice that we have not applied the  $\mathcal{O}(1)$ initial condition. This is done later on. We note that (\ref{46}) is equivalent to equation (\ref{TE9}) for $\mathbf{K}_{0}(t_{1},t_{2})$ using (\ref{Conexion}).

We now solve the $\mathcal{O}(\epsilon)$ problem in (\ref{45}). Substituting $\mathbf{Y}_{0}(t_{1},t_{2})$ given in (\ref{46}) into the $\mathcal{O}(\epsilon)$ differential equation in (\ref{45}) and solving the resulting equation one finds that 
\begin{eqnarray}
\label{47}
\mathbf{Y}_{1}(t_{1},t_{2}) &=& \mathbf{Y}_{1}(0,t_{2}) + \left[ \mathbb{A}_{0}\mathbf{Y}_{0}(0,t_{2}) - \frac{\partial \mathbf{Y}_{0}}{\partial t_{2}}(0,t_{2}) \right] t_{1} \cr
&& + \left[ \mathbb{A}_{2}(t_{1}) - \mathbb{A}_{2}(0) \right] \mathbf{Y}_{0}(0,t_{2}) \ ,
\end{eqnarray}
where $\mathbb{A}_{0}$ and $\mathbb{A}_{2}(t_{1})$ are given in (\ref{DefinicionMatrices}).

Now one must eliminate \textit{secular terms} from $\mathbf{Y}_{1}(t_{1},t_{2})$. These correspond to terms that destroy the order of the asymptotic expansion in (\ref{44}) because they  become unbounded as $t_{1}$ increases. From (\ref{47}) it follows that secular terms disappear from $\mathbf{Y}_{1}(t_{1},t_{2})$ if and only if the term in brackets multiplied by $t_{1}$ is zero for all $t_{2}$, that is, if and only if 
\begin{eqnarray}
\label{48ii}
\mathbb{A}_{0}\mathbf{Y}_{0}(0,t_{2}) - \frac{\partial \mathbf{Y}_{0}}{\partial t_{2}}(0,t_{2}) = 0 \ .
\end{eqnarray}
We note that (\ref{48ii}) is equivalent to equation (\ref{TE10}) for $(\partial \mathbf{K}_{0}/\partial t_{2})(0,t_{2})$ using (\ref{Conexion}).

Notice that (\ref{48ii}) is identical to the system of equations in the RWA given in (\ref{RWAeq}) if one makes the changes 
\begin{eqnarray}
\label{cambios}
d/d\tau \rightarrow \partial/\partial t_{2} \ , \quad \epsilon \rightarrow 1 \ , \quad \tau \rightarrow t_{2} \ , \quad \mathbf{X}(\tau) \rightarrow \mathbf{Y}_{0}(0,t_{2}) . \cr
&&
\end{eqnarray}
Hence, one can use the result in (\ref{RWA})  to determine the solution of (\ref{48ii}). One obtains
\begin{eqnarray}
\label{57}
\mathbf{Y}_{0}(t_{1},t_{2}) = \mathbf{Y}_{0}(0,t_{2}) = e^{\mathbb{A}_{0}t_{2}}\mathbf{X}(0) = \mathbb{Q} e^{\mathbb{D} t_{2}}\mathbb{Q}^{-1} \mathbf{X}(0)  \ , \cr
&&
\end{eqnarray}
where $\mathbb{Q}$ and $e^{\mathbb{D}t_{2}}$ are given in (\ref{DefinicionMatrices}). Notice that in writing (\ref{57}) we applied the initial condition for $\mathbf{Y}_{0}(t_{1},t_{2})$ given in the $\mathcal{O}(1)$ problem in (\ref{45}) and we used the result of the $\mathcal{O}(1)$ equation in (\ref{46}). Also, without secular terms $\mathbf{Y}_{1}(t_{1},t_{2})$ reduces to 
\begin{eqnarray}
\label{49}
\mathbf{Y}_{1}(t_{1},t_{2}) &=& \mathbf{Y}_{1}(0,t_{2}) + \left[ \mathbb{A}_{2}(t_{1}) - \mathbb{A}_{2}(0) \right] \mathbf{Y}_{0}(0,t_{2}) \ , \cr
&&
\end{eqnarray}
Notice that we have not applied the $\mathcal{O}(\epsilon)$ initial condition. This is done later on. 

We now solve the $\mathcal{O}(\epsilon^{2})$ problem in (\ref{45}). Substituting the expression for $\mathbf{Y}_{1}(t_{1},t_{2})$ in (\ref{49}) into the $\mathcal{O}(\epsilon^{2})$ differential equation and solving the resulting equation one obtains that
\begin{widetext}
\begin{eqnarray}
\label{52}
\mathbf{Y}_{2}(t_{1},t_{2}) &=& \mathbf{Y}_{2}(0,t_{2}) + \left[ \mathbb{A}_{2}(t_{1}) - \mathbb{A}_{2}(0) \right]\mathbf{Y}_{1}(0,t_{2})  + \Big\{ \  [\mathbb{A}_{0}, \mathbb{A}_{1}(0) - \mathbb{A}_{1}(t_{1}) ] + \mathbb{A}_{4}(t_{1}) - \mathbb{A}_{4}(0)- \mathbb{A}_{2}(t_{1})\mathbb{A}_{2}(0)   \cr
&&  + \mathbb{A}_{2}(0)^{2} \ \Big\} \mathbf{Y}_{0}(0,t_{2}) -t_{1} \left\{ \frac{\partial \mathbf{Y}_{1}}{\partial t_{2}} (0,t_{2}) - \mathbb{A}_{0}\mathbf{Y}_{1}(0,t_{2}) + \Big( \left[ \mathbb{A}_{0}, \mathbb{A}_{2}(0) \right] - \mathbb{A}_{3} \Big) \mathbf{Y}_{0}(0,t_{2})  \right\} \cr
&&
\end{eqnarray}
\end{widetext}
where all the matrices are defined in (\ref{DefinicionMatrices}).

Now we have to eliminate secular terms from $\mathbf{Y}_{2}(t_{1},t_{2})$. Again, in our case these are terms that destroy de order of the asymptotic expansion in (\ref{44}) because they become unbounded as $t_{1}$ increases. From (\ref{52}) it follows that secular terms disappear from $\mathbf{Y}_{2}(t_{1},t_{2})$ if and only if the term in parenthesis multiplied by $t_{1}$ is zero for all $t_{2}$, that is, if and only if 
\begin{eqnarray}
\label{XXX}
\frac{\partial \mathbf{Y}_{1}}{\partial t_{2}} (0,t_{2}) = \mathbb{A}_{0}\mathbf{Y}_{1}(0,t_{2}) - \Big( \left[ \mathbb{A}_{0}, \mathbb{A}_{2}(0) \right] - \mathbb{A}_{3} \Big) \mathbf{Y}_{0}(0,t_{2})  \ . \cr
&&
\end{eqnarray}
We note that (\ref{XXX}) is equivalent to equation (\ref{TE13}) for $(\partial \mathbf{K}_{1} / \partial t_{2} ) (0,t_{2})$ using (\ref{Conexion}).

Solving (\ref{XXX}) and applying the $\mathcal{O}(\epsilon)$ initial condition for $\mathbf{Y}_{1}(t_{1},t_{2})$ given in (\ref{45}) leads to the result
\begin{eqnarray}
\label{62}
\mathbf{Y}_{1}(0,t_{2}) = \mathbb{G}_{1}(t_{2})\mathbf{X}(0) ,
\end{eqnarray}
where $\mathbb{G}_{1}(t_{2})$ is defined in (\ref{DefinicionMatrices}). If one substitutes this result into the expression for $\mathbf{Y}_{1}(t_{1},t_{2})$ given in (\ref{49}) one obtains that
\begin{eqnarray}
\label{63}
\mathbf{Y}_{1}(t_{1},t_{2}) = \mathbb{W}(t_{1},t_{2}) e^{\mathbb{A}_{0}t_{2}} \mathbf{X}(0) ,
\end{eqnarray}
where $ \mathbb{W}(t_{1},t_{2})$ and $\mathbb{A}_{0}$ are defined in (\ref{DefinicionMatrices}).

Before proceeding we observe that, without secular terms, $\mathbf{Y}_{2}(t_{1},t_{2})$ in (\ref{52}) reduces to
\begin{widetext}
\begin{eqnarray}
\label{54}
\mathbf{Y}_{2}(t_{1},t_{2}) &=& \mathbf{Y}_{2}(0,t_{2}) + \left[ \mathbb{A}_{2}(t_{1}) - \mathbb{A}_{2}(0) \right]\mathbf{Y}_{1}(0,t_{2})  + \Big\{ \  [\mathbb{A}_{0}, \mathbb{A}_{1}(0) - \mathbb{A}_{1}(t_{1}) ] + \mathbb{A}_{4}(t_{1}) - \mathbb{A}_{4}(0)- \mathbb{A}_{2}(t_{1})\mathbb{A}_{2}(0)  \cr
&& + \mathbb{A}_{2}(0)^{2} \ \Big\} \mathbf{Y}_{0}(0,t_{2}) .
\end{eqnarray}
\end{widetext}
Notice that the $\mathcal{O}(\epsilon^{2})$ initial condition for $\mathbf{Y}_{2}(t_{1},t_{2})$ in (\ref{45}) has not yet been applied. This initial condition would be used when eliminating secular terms for $\mathbf{Y}_{3}(t_{1},t_{2})$ after solving the $\mathcal{O}(\epsilon^{3})$ differential equation.

\subsection{The one- and two-term approximations}

Since $\mathbf{Y}_{0}(t_{1},t_{2})$ and $\mathbf{Y}_{1}(t_{1},t_{2})$ are now completely specified, one can write a one- and two-term approximations for $\mathbf{X}(\tau)$. This is what we do now.

First, we consider the one-term approximation. From the asymptotic expansion in (\ref{44}) and the expression for $\mathbf{Y}_{0}(t_{1},t_{2})$ given in (\ref{57}) it follows that a one-term approximation of $\mathbf{Y}(t_{1},t_{2})$ is given by 
\begin{eqnarray}
\label{64}
\mathbf{Y}(t_{1},t_{2}) &\simeq& \mathbf{Y}_{0}(t_{1},t_{2}) = e^{\mathbb{A}_{0}t_{2}} \mathbf{X}(0) .
\end{eqnarray} 
We note that this result is equivalent to that for $\mathbf{K}(t_{1},t_{2})$ in (\ref{Nuevo1termino}) using (\ref{ConexionYZ}) and (\ref{Conexion}).

Substituting (\ref{64}) in (\ref{41}) and using the definition of the two time-scales $t_{1}$ and $t_{2}$ in (\ref{40}) one obtains a one-term approximation for $\mathbf{X}(\tau)$:
\begin{eqnarray}
\label{x0}
\mathbf{X}(\tau) \simeq \mathbf{X}_{0}(\tau) \equiv e^{\epsilon \mathbb{A}_{0}\tau} \mathbf{X}(0) \ .
\end{eqnarray}
Comparing (\ref{RWA}) with (\ref{x0}) one concludes that the one-term approximation $\mathbf{X}_{0}(\tau)$ to $\mathbf{X}(\tau)$ corresponds to the solution in the RWA. 

To obtain an explicit expression for the one-term approximation one simply substitutes in (\ref{x0}) the factorization of $e^{\mathbb{A}_{0}t_{2}}$ given in (\ref{factorizacion}) and carries out the multiplication of matrices. The result is \begin{eqnarray}
\label{Aprox1termino}
\rho_{12}(\tau) &\simeq& \rho_{12}(0)\mbox{cos}^{2}\left( \frac{\epsilon\tau}{2} \right) + \rho_{12}(0)^{*}\mbox{sin}^{2}\left( \frac{\epsilon\tau}{2} \right) \cr
&& + \frac{i}{2} \alpha_{30}(0)\mbox{sin}\left( \epsilon\tau \right) \ , \cr
&& \cr
\rho_{21}(\tau) &=& \rho_{12}(\tau)^{*} \ , \cr
&& \cr
\alpha_{30}(\tau) &\simeq& i\rho_{12}(0) \mbox{sin}\left( \epsilon\tau \right) - i\rho_{12}(0)^{*}\mbox{sin} \left(  \epsilon\tau \right) \cr
&& + \alpha_{30}(0)\mbox{cos}\left( \epsilon\tau \right) \ \ .
\end{eqnarray}
Introducing units in (\ref{Aprox1termino}) using (\ref{16}), one obtains the results in (\ref{RWAsolucionE}).

Now, we consider the two-term approximation.  From the asymptotic expansion in (\ref{44}) and the expressions for $\mathbf{Y}_{0}(t_{1},t_{2})$ and $\mathbf{Y}_{1}(t_{1},t_{2})$ given in (\ref{57}) and (\ref{63}), respectively, it follows that a two-term approximation of $\mathbf{Y}(t_{1},t_{2})$ is given by 
\begin{eqnarray}
\label{64b}
\mathbf{Y}(t_{1},t_{2}) &\simeq& \mathbf{Y}_{0}(t_{1},t_{2}) + \epsilon \mathbf{Y}_{1}(t_{1},t_{2})  \cr
&=& \left[ \mathbb{I}_{3} + \epsilon \mathbb{W}(t_{1},t_{2}) \right] e^{\mathbb{A}_{0}t_{2}} \mathbf{X}(0) . 
\end{eqnarray} 
Here $\mathbb{I}_{3}$ is the $3 \times 3$ identity matrix. We note that this result is equivalent to that for $\mathbf{K}(t_{1},t_{2})$ in (\ref{Nuevo2termino}) using (\ref{ConexionYZ}) and (\ref{Conexion}).

Substituting (\ref{64b}) in (\ref{41}) and using the definition of the two time-scales $t_{1}$ and $t_{2}$ in (\ref{40}) one obtains a two-term approximation for $\mathbf{X}(\tau)$:
\begin{eqnarray}
\label{x1}
\mathbf{X}(\tau) \simeq \mathbf{X}_{1}(\tau) \equiv \left[ \mathbb{I}_{3} + \epsilon \mathbb{W}(\tau, \epsilon\tau ) \right] e^{\epsilon \mathbb{A}_{0} \tau} \mathbf{X}(0)  \ .
\end{eqnarray}
Carrying out the multiplication of matrices on the righthand side of (\ref{x1}) one obtains an explicit expression for the two-term approximation
\begin{widetext}
\begin{eqnarray}
\label{Aprox2termino}
\rho_{12}(\tau) &\simeq& \rho_{12}(0)\left[ \mbox{cos}^{2}\left( \frac{\epsilon\tau}{2} \right) - i\frac{\epsilon}{2}e^{-i\tau}\mbox{sin}(\epsilon \tau) \right] + \rho_{12}(0)^{*}\left[ \mbox{sin}^{2}\left( \frac{\epsilon\tau}{2} \right) -i\frac{\epsilon}{2}\mbox{sin}(\epsilon \tau) \left( 1 - e^{-i\tau} \right) \right] \cr
&& + \alpha_{30}(0) \left\{  \frac{i}{2} \mbox{sin}\left( \epsilon\tau \right) + \frac{\epsilon }{2} \left[ \mbox{cos}^{2}\left( \frac{\epsilon \tau}{2} \right) - e^{-i\tau}\mbox{cos}(\epsilon \tau)  \right] \right\} \ , \cr
&& \cr
&& \cr
\alpha_{30}(\tau) &\simeq& \rho_{12}(0) \left( \ i \mbox{sin}\left( \epsilon\tau \right)  -\frac{\epsilon}{2}\Big\{ 1 -2\mbox{cos}(\tau) + \mbox{cos}(\epsilon \tau) \left[ 1 -i2\mbox{sin}(\tau) \right]  \Big\} \right) \cr
&&
 + \rho_{12}(0)^{*} \left( \ -i \mbox{sin}\left( \epsilon\tau \right) -\frac{\epsilon}{2}\Big\{ 1 -2\mbox{cos}(\tau) + \mbox{cos}(\epsilon \tau) \left[ 1 +i2\mbox{sin}(\tau) \right]  \Big\} \right) \cr
&& + \alpha_{30}(0)\left[ \ \mbox{cos}\left( \epsilon\tau \right) - \epsilon \mbox{sin}(\epsilon \tau)\mbox{sin}(\tau) \ \right] \ .
\end{eqnarray}
\end{widetext}
Observe that $\alpha_{30}(\tau)$ is indeed a real quantity. Also, the two-term approximation does satisfy $\rho_{21}(\tau) = \rho_{12}(\tau)^{*}$. Introducing units in (\ref{Aprox2termino}) using (\ref{16}), one obtains the results in (\ref{Aprox2terminoU}).

Now the question arises as to whether (\ref{Aprox2termino}) gives rise to a density operator or not. Now, the one-term approximation in (\ref{Aprox1termino}) does lead to a density operator because it is exactly equal to the solution in the RWA and the solution in RWA corresponds to solving von Neumann's equation with a Hermitian Hamiltonian, see either Section III or Appendix A.2. Nevertheless, the same cannot be said about the two-term approximation.

First of all, the matrix representation of $\tilde{\rho}_{I}(\tau)$ in the basis $\gamma = \{  \ \vert 1 \rangle , \ \vert 2 \rangle  \ \}$ is 
\begin{eqnarray}
\label{65}
\left[ \tilde{\rho}_{I}(\tau) \right]_{\gamma} = \left(
\begin{array}{cc}
\frac{1}{2}\left[ 1  - \alpha_{30}(\tau) \right] & \rho_{12}(\tau) \cr
\rho_{12}(\tau)^{*} &  \frac{1}{2}\left[ 1  + \alpha_{30}(\tau) \right]
\end{array}
\right)  .
\end{eqnarray}
Observe that we used in (\ref{65}) the definitions in (\ref{16}) and the following three results:
\begin{eqnarray}
\label{65b}
\rho_{11}(\tau) &=& 1 - \rho_{22}(\tau) \ , \quad \rho_{21}(\tau) = \rho_{12}(\tau)^{*} \ , \cr 
\rho_{22}(\tau) &=& \frac{1}{2}\left[ 1  + \alpha_{30}(\tau) \right] .
\end{eqnarray}
The first two follow from the fact that $\tilde{\rho}_{I}(\tau)$ is a Hermitian operator with trace equal to $1$ because $\tilde{\rho}_{I}(\tau)$ is a density operator. The third follows from the definition of $\alpha_{30}(\tau)$ in (\ref{16}) and the first equation in (\ref{65b}) expressing $\rho_{11}(\tau)$ in terms of $\rho_{22}(\tau)$. Now, $\left[ \tilde{\rho}_{I}(\tau) \right]_{\gamma}$ is a density matrix if and only if it is Hermitian and positive semidefinite with trace equal to one. Equivalently, $\left[ \tilde{\rho}_{I}(\tau) \right]_{\gamma}$ is a density matrix if and only if it is Hermitian with trace equal to one and with nonnegative eigenvalues. The eigenvalues of  $\left[ \tilde{\rho}_{I}(\tau) \right]_{\gamma}$ are easily calculated to be
\begin{eqnarray}
\label{66}
p_{\pm}(\tau) =  \frac{1}{2}\left\{ \ 1 \pm \sqrt{ \alpha_{30}(\tau)^{2} + 4\vert \rho_{12}(\tau) \vert^{2} } \  \right\} \ .
\end{eqnarray}
If one substitutes into (\ref{65}) and (\ref{66}) the two-term approximation given in the righthand side of (\ref{Aprox2termino}), then it is straightforward to show that one obtains a Hermitian matrix that has trace equal to $1$ and whose eigenvalues are
\begin{eqnarray}
\label{155}
p_{\pm}(\tau) =  \frac{1}{2}\left\{ \ 1 \pm \sqrt{ \left\vert \mathbf{r}_{I}(0) \right\vert^{2} +  \epsilon^{2}\left[ \left\vert E_{1}(\tau)  \right\vert^{2} + \left\vert E_{2}(\tau)  \right\vert^{2} \right] } \  \right\} , \cr
\end{eqnarray}
where 
\begin{eqnarray}
\label{156}
\mathbf{r}_{I}(0) \ = \ \left( \ 2\mbox{Re}\left[ \rho_{12}(0) \right] , \ 2\mbox{Im}\left[ \rho_{12}(0) \right] , \ \alpha_{30}(0) \ \right) \ , \cr
\end{eqnarray}
and 
\begin{eqnarray}
\label{151}
E_{1}(\tau) &=& \mbox{sin}(\epsilon \tau) \left( 2e^{-i\tau} - 1 \right) \mbox{Im}\left[ \rho_{12}(0) \right] \cr
&& - i\mbox{sin}(\epsilon \tau) \mbox{Re}\left[ \rho_{12}(0) \right]  \cr
&& + \alpha_{30}(0) \left[ \mbox{cos}^{2}\left( \frac{\epsilon \tau}{2} \right) - e^{-i\tau}\mbox{cos}(\epsilon \tau) \right] \ , \cr
E_{2} (\tau) &=& 2\left[ \mbox{cos}^{2}\left( \frac{\epsilon \tau}{2} \right) - \mbox{cos}(\tau) \right] \mbox{Re}\left[ \rho_{12}(0) \right] \cr
&& + \mbox{sin}(\epsilon \tau) \mbox{sin}(\tau)\alpha_{30}(0) \cr
&& + 2\mbox{cos}(\epsilon \tau) \mbox{sin} (\tau) \mbox{Im}\left[ \rho_{12}(0) \right] \ .
\end{eqnarray}
Here $\mathbf{r}_{I}(0)$ is the Bloch vector associated with $\tilde{\rho}_{I}(0) = \rho_{I}(0) = \rho(0)$ (recall from equations (\ref{11}) and (\ref{12}) that the Sch\"{o}dinger and Interaction pictures coincide at $t=0$ and from (\ref{16}) that $\tilde{\rho}_{I}(\tau) = \rho_{I}[\tau/(2\omega_{1})]$).

From (\ref{155}) it is clear that $p_{+}(\tau) >0$ and that $p_{-}(\tau)$ could be negative. As a consequence, one may not obtain a density matrix  if one substitutes into (\ref{65}) the two-term approximation given in the righthand side of (\ref{Aprox2termino}). For example, if $\tilde{\rho}_{I}(0) = \rho_{I}(0) = \rho(0)$ is a pure state, then $\vert \mathbf{r}_{I}(0) \vert = 1$ (see item (i) after (\ref{ComponentesBloch})) and, in general, $p_{-}(\tau) < 0$. Therefore, one does not obtain a density matrix if $\rho (0)$ is a pure state. One can remedy this difficulty by proceeding as described in Section IV.

Before ending this section we give a geometrical interpretation of the above difficulty. The Bloch vector $\mathbf{r}_{I}(\tau)$ is defined by
\begin{eqnarray}
\label{BlochA}
\mathbf{r}_{I}(\tau) &=& \left( \ \alpha_{10}(\tau) , \alpha_{20}(\tau) , \alpha_{30}(\tau) \ \right) \ ,
\end{eqnarray}
where 
\begin{eqnarray}
\label{BlochA2}
\alpha_{10}(\tau) &=& 2\mbox{Re}\left[ \rho_{12}(\tau)  \right] \ , \quad  \alpha_{20}(\tau) \ = \ 2\mbox{Im}\left[ \rho_{12}(\tau)  \right] \ , \cr
&&
\end{eqnarray}
and $\alpha_{30}(\tau)$ and $\rho_{12}(\tau)$ are given in (\ref{16}). Notice that $\mathbf{r}_{I}(\tau)$ in (\ref{BlochA}) is identical to $\mathbf{r}_{I}(t)$ in (\ref{Bloch}), except that the latter is a function of $t = \tau/(2\omega_{1})$, while the former is a function of the non dimensional time $\tau$. 

Substituting into (\ref{BlochA}) and (\ref{BlochA2}) the two-term approximation given in the righthand side of (\ref{Aprox2termino}) one obtains an approximate Bloch vector $\mathbf{r}_{I}(\tau)$ whose (Euclidean) norm is given by
\begin{eqnarray}
\label{NormaBloch}
\vert \mathbf{r}_{I}(\tau) \vert &=&  \sqrt{ \alpha_{30}(\tau)^{2} + 4\vert \rho_{12}(\tau) \vert^{2} } \  ,  \cr 
&=& \sqrt{ \left\vert \mathbf{r}_{I}(0) \right\vert^{2} +  \epsilon^{2}\left[ \left\vert E_{1}(\tau)  \right\vert^{2} + \left\vert E_{2}(\tau)  \right\vert^{2} \right] }  \ , \cr
&&
\end{eqnarray}
see the terms inside the square roots in equations (\ref{66}) and (\ref{155}). Therefore, the two term approximation increases slightly the length of the initial Bloch vector $\mathbf{r}_{I}(0)$. As mentioned above, this can be remedied by proceeding as described in Section IV.

\subsection{The numerical calculation}

In this section we present how the numerical results were obtained. We used the Bloch vector defined in (\ref{BlochA}) and (\ref{BlochA2}). First, observe that, if $\mathbf{r}_{I}(0)$ is a pure-state, then $\vert \mathbf{r}_{I}(0) \vert = 1$ and, thus, $\mathbf{r}_{I}(0)$ is located on the surface of the sphere centered at the origin and with radius $1$ (the Bloch sphere). Hence, $\mathbf{r}_{I}(0)$ can be expressed in spherical coordinates as follows:
\begin{eqnarray}
\label{BlochA3}
\mathbf{r}_{I}(0) &=& \Big( \ \mbox{sin}(\Theta)\mbox{cos}(\Phi) , \mbox{sin}(\Theta)\mbox{sin}(\Phi) , \mbox{cos}(\Theta) \Big) 
\end{eqnarray}
where $\Theta$ is the polar angle, while $\Phi$ is the azimuthal angle.

Second, we determined the differential equations governing the evolution of the components of $\mathbf{r}_{I}(\tau)$. From the system of differential equations in (\ref{25}) with the matrices defined in (\ref{DefinicionMatrices}) and the definition of the components of the Bloch vector in (\ref{BlochA2}) it is straightforward to show that
\begin{eqnarray}
\label{eqBloch}
\frac{d}{d\tau} \alpha_{10}(\tau) &=& \epsilon\mbox{sin}(\tau) \alpha_{30}(\tau) \ , \cr
\frac{d}{d\tau} \alpha_{20}(\tau) &=& \epsilon\left[ 1+ \mbox{cos}(\tau) \right] \alpha_{30}(\tau) \ , \cr
\frac{d}{d\tau} \alpha_{30}(\tau) &=& -\epsilon\Big\{ \ \mbox{sin}(\tau) \alpha_{10}(\tau) + \left[ 1 +\mbox{cos}(\tau) \right] \alpha_{20}(\tau) \ \Big\} \  . \cr
&&
\end{eqnarray}
These are called the \textit{optical Bloch equations} and they are equivalent to  the system of differential equations in (\ref{25}).

Then, we defined three meshes:
\begin{enumerate}
\item The $\epsilon$-mesh: It is a uniform mesh from $\epsilon = \epsilon_{min}$ to $\epsilon = \epsilon_{max}$ with step length $\Delta \epsilon = 0.0025$. In Figure \ref{Figure1a} $\epsilon_{min} = 1/20 = 0.05$ and $\epsilon_{max} = 1/4 = 0.25$, while in Figure \ref{Figure1b} $\epsilon_{min} = 1/50 = 0.02$ and $\epsilon_{max} = 1/8 = 0.125$.

\item The $\Theta\Phi$-mesh: It is the $2$ dimensional set 
\begin{eqnarray}
\label{MallaE}
&& \Big\{ (\Theta_{j}, \ \Phi_{k}) = (0.1j , \  0.1k) :  \ j = 1,2,...,31 \ , \cr
&&  \quad\quad\quad\quad\quad\quad\quad\quad\quad\quad\quad\quad k =0,1,...,62 \Big\} \cr
&\cup& \left\{  \ (\Theta_{0}, \ \Phi_{0}) = (0,0) , \ (\Theta_{32}, \ \Phi_{0}) = (\pi , 0) \ \right\}  .
\end{eqnarray}
Here $\Theta_{j}$ is a polar angle, while $\Phi_{k}$ is an azimuthal angle. Notice that, in the first set in (\ref{MallaE}), $\Theta_{j}$ ranges from $0.1$ to $3.1$ with uniform step length $\Delta\Theta = 0.1$ and $\Phi_{k}$ ranges from $0$ to $6.2$ with uniform step length $\Delta\Phi = 0.1$. Also, observe that $\Delta\Theta = \Delta\Phi = 0.1$ radians is equal to $5.73$ degrees (this is a bit less than one minute in an analogue watch). This mesh was introduced to construct any initial Bloch vector on the surface of the Bloch sphere according to (\ref{BlochA3}). The second set in (\ref{MallaE}) was introduced to take into account the north and south poles of the surface of the Bloch sphere. 

\item The $\tau$-mesh: It is a uniform mesh from $\tau = 0$ to $\tau = \tau_{max}$ with uniform step length $\Delta \tau = 0.001$. Also, $\tau_{max} = 2\pi/\epsilon$ in Figure \ref{Figure1a} and $\tau_{max} = 10(2\pi/\epsilon)$ in Figure \ref{Figure1b}. Observe that  $2\pi/\epsilon$ corresponds to one Rabi oscillation, so that the $\tau$-mesh goes from $0$ to one Rabi oscillation in Figure \ref{Figure1a} and from $0$ to $10$ Rabi oscillations in Figure \ref{Figure1b}. Notice that one has a different $\tau$-mesh for each value of $\epsilon$.
\end{enumerate}
After defining the meshes, for each value of $\epsilon$ in the $\epsilon$-mesh we proceeded in the following steps: 
\begin{enumerate}
\item For each value $(\Theta_{j}, \ \Phi_{k})$ in the $\Theta\Phi$-mesh we constructed the initial Bloch vector
\begin{eqnarray}
&& \mathbf{r}_{I}(0;\Theta_{j}, \ \Phi_{k}) \cr
&=& \left( \ \mbox{sin}(\Theta_{j})\mbox{cos}(\Phi_{k}) , \ \mbox{sin}(\Theta_{j})\mbox{sin}(\Phi_{k}) , \ \mbox{cos}(\Theta_{j}) \ \right) \ . \cr
&&
\end{eqnarray}
Notice that $\mathbf{r}_{I}(0;\Theta_{j}, \ \Phi_{k})$ is located on the surface of the Bloch sphere, so that it corresponds to a pure state.

\item For each value $(\Theta_{j}, \ \Phi_{k})$ in the $\Theta\Phi$-mesh we took $\mathbf{r}_{I}(0;\Theta_{j}, \ \Phi_{k})$ as the initial condition and we calculated four solutions:
\begin{enumerate}
\item A numerical solution $\mathbf{r}_{num}(\tau;\Theta_{j}, \ \Phi_{k}) $ of the Bloch vector by numerically solving the optical Bloch equations in (\ref{eqBloch}).
\item A two-term approximate solution $\mathbf{r}_{I}(\tau;\Theta_{j}, \ \Phi_{k})$ of the Bloch vector by using (\ref{Aprox2termino}),  (\ref{BlochA}), and (\ref{BlochA2})
\item A two-term normalized approximate solution of the Bloch vector
\[
\mathbf{r}_{IN}(\tau;\Theta_{j}, \ \Phi_{k}) = \frac{\mathbf{r}_{I}(\tau;\Theta_{j}, \ \Phi_{k})}{\vert \mathbf{r}_{I}(\tau;\Theta_{j}, \ \Phi_{k}) \vert} \  .
\]
\item A Bloch vector $\mathbf{r}_{RWA}(\tau;\Theta_{j}, \ \Phi_{k})$ in the RWA by using using (\ref{Aprox1termino}), (\ref{BlochA}), and (\ref{BlochA2}).
\end{enumerate}

\item For each value $(\Theta_{j}, \ \Phi_{k})$ in the $\Theta\Phi$-mesh we took $\mathbf{r}_{num}(\tau;\Theta_{j}, \ \Phi_{k})$ as the exact solution and we calculated the relative error of each of the approximate Bloch vectors in the $\tau$-mesh:
\begin{widetext}
\begin{eqnarray}
\label{errores1}
e_{R}(\tau ; \Theta_{j}, \ \Phi_{k}) &=& \frac{\left\vert \mathbf{r}_{num}(\tau;\Theta_{j}, \ \Phi_{k}) -  \mathbf{r}_{I}(\tau;\Theta_{j}, \ \Phi_{k}) \right\vert}{\left\vert \mathbf{r}_{num}(\tau;\Theta_{j}, \ \Phi_{k}) \right\vert} \ , \quad e_{RN}(\tau ; \Theta_{j}, \ \Phi_{k}) = \frac{\left\vert \mathbf{r}_{num}(\tau;\Theta_{j}, \ \Phi_{k}) -  \mathbf{r}_{IN}(\tau;\Theta_{j}, \ \Phi_{k}) \right\vert}{\left\vert \mathbf{r}_{num}(\tau;\Theta_{j}, \ \Phi_{k}) \right\vert} \  , \cr
&& \cr
e_{R}^{RWA}(\tau ; \Theta_{j}, \ \Phi_{k}) &=& \frac{\left\vert \mathbf{r}_{num}(\tau;\Theta_{j}, \ \Phi_{k}) -  \mathbf{r}_{RWA}(\tau;\Theta_{j}, \ \Phi_{k}) \right\vert}{\left\vert \mathbf{r}_{num}(\tau;\Theta_{j}, \ \Phi_{k}) \right\vert} \ .
\end{eqnarray}
\end{widetext}
Recall that  $\mathbf{r}_{num}(\tau;\Theta_{j}, \ \Phi_{k}) $ corresponds to a pure-state, so that $\vert \mathbf{r}_{num}(\tau;\Theta_{j}, \ \Phi_{k}) \vert = 1$. For this reason we did not divide each of the numerators in the equations in (\ref{errores1}) by this quantity.

\item We took the maximum value of each of the relative errors in the $\Theta\Phi$- and $\tau$-meshes:
\begin{widetext}
\begin{eqnarray}
\label{errores2}
E_{R} &=& \mbox{max}\Big\{ e_{R}(\tau ; \Theta_{j}, \ \Phi_{k}) : \ \ \mbox{$\tau$ is in the $\tau$-mesh,  $(\Theta_{j},\Phi_{k})$ is in the $\Theta\Phi$-mesh}   \Big\} \ ,\cr
E_{RN} &=& \mbox{max}\Big\{ e_{RN}(\tau ; \Theta_{j}, \ \Phi_{k}) : \ \ \mbox{$\tau$ is in the $\tau$-mesh, $(\Theta_{j},\Phi_{k})$ is in the $\Theta\Phi$-mesh}   \Big\} \ ,\cr
E_{R}^{RWA} &=& \mbox{max}\Big\{ e_{R}^{RWA}(\tau ; \Theta_{j}, \ \Phi_{k}) : \ \ \mbox{$\tau$ is in the $\tau$-mesh, $(\Theta_{j},\Phi_{k})$ is in the $\Theta\Phi$-mesh}   \Big\} \ .
\end{eqnarray}
\end{widetext}
We emphasize that $E_{R}$ is the maximum relative error that one can have if one uses the two-term approximation in (\ref{Aprox2termino}) from $\tau = 0$ to $\tau = \tau_{max}$ with any initial condition that corresponds to a pure state; $E_{RN}$ is the maximum relative error that one can have if one uses the normalized two-term approximation from $\tau = 0$ to $\tau = \tau_{max}$ with any initial condition that corresponds to a pure state; and $E_{R}^{RWA}$ is the maximum relative error that one can have if one uses the approximation in the RWA given in (\ref{Aprox1termino}) from $\tau = 0$ to $\tau = \tau_{max}$ with any initial condition that corresponds to a pure state. 
\end{enumerate}
Therefore, for each value of $\epsilon$ from $\epsilon_{min}$ to $\epsilon_{max}$ we have the values of $E_{R}$, $E_{RN}$, and $E_{R}^{RWA}$. Each of the sets of data $(\epsilon , E_{R})$, $(\epsilon , E_{RN})$, and $(\epsilon , E_{R}^{RWA})$ was then interpolated using cubic splines with a not-a-knot-condition \cite{Burden} between $\epsilon_{min}$ and $\epsilon_{max}$  to produce the results in Figures \ref{Figure1a} and \ref{Figure1b}.

%%%%%%%%%%%%%%
%%%%%%%%%%%%%%
%%%%%%%%%%%%%%
%%%%%%%%%%%%%%


\begin{thebibliography}{99}

\bibitem{80} D. Braak et al, J. Phys. A: Math. Theor. \textbf{49}, 300301 (2016).

\bibitem{libro} S. Haroche and J. M. Raimond, \textit{Exploring the Quantum: Atoms, Cavities, and Photons} (Oxford University Press, 2006).

\bibitem{Maser1} H. Walther, B. T. H. Varcoe, B. G. Englert, and T. Becker, Rep. Prog. Phys. \textbf{69}, 1325 (2006).

\bibitem{Maser2} J. M. Raimond, M. Brune, and S. Haroche, Reviews of Modern Physics \textbf{73}, 565 (2001).

\bibitem{Shakov} Kh. Kh. Shakov and J. H. McGuire, Phys. Rev. A \textbf{67}, 033405 (2003).

\bibitem{Devoret} M. Devoret, S. Girvin, and R. Schoelkopf, Ann. Phys. \textbf{16}, 767 (2007).

\bibitem{Niem}  T. Niemczyk et al., Nat. Phys. \textbf{6}, 772 (2010).

\bibitem{Forn}  P. Forn-D\'{i}az, J. Lisenfeld, D. Marcos, J. J. Garc\'{i}a-Ripoll, E. Solano, C. J. P. M. Harmans, and J. E. Mooij, Phys. Rev. Lett. \textbf{105}, 237001 (2010).

\bibitem{Ion} J. S. Pedernales, I. Lizuain, S. Felicetti, G. Romero, L. Lamata, and E. Solano, Sci. Rep. \textbf{5}, 15472 (2015).

\bibitem{Braak} D. Braak, Phys. Rev. Lett. \textbf{107}, 100401 (2011).

\bibitem{Nuevo1} Q. H. Chen, C. Wang, S. He, T. Liu, and K. L. Wang, Phys. Rev. A \textbf{86}, 023822 (2012).

\bibitem{Nuevo2} H. H. Zhong, Q. T. Xie, M. T. Batchelor, and C. H. Lee, J. Phys. A \textbf{46}, 415302 (2013).

\bibitem{Nuevo3}  A. J. Maciejewski, M. Przybylska, and T. Stachowiak, Phys. Lett. A \textbf{378}, 3445 (2014).

\bibitem{IrishPRB} E. K. Irish, J. Gea-Banacloche, I. Martin, and K. C. Schwab, Phys. Rev. B \textbf{72}, 195410 (2005).

\bibitem{Irish}  E. K. Irish, Phys. Rev. Lett. \textbf{99}, 173601 (2007).

\bibitem{Hausinger} J. Hausinger and M. Grifoni, Phys Rev A \textbf{82}, 062320 (2010).

\bibitem{Zhang} Y. Y. Zhang and Q. H. Chen, Phys. Rev. A \textbf{91}, 013814 (2015).

\bibitem{Casanova} J. Casanova, G. Romero, I. Lizuain, J. J. Garc\'{i}a-Ripoll, and E. Solano, Phys. Rev. Lett. \textbf{105}, 263603 (2010).

\bibitem{Ashab} S. Ashhab and F. Nori, Phys. Rev. A \textbf{81}, 042311 (2010).

\bibitem{Eberly} S. Agarwal, S. M. Hashemi Rafsanjani, and J. H. Eberly, Phy. Rev. A \textbf{85}, 043815 (2012). 

\bibitem{Zhang2} Y. Y. Zhang, X. Y. Chen, S. He, and Q. H. Chen, Phys. Rev. A \textbf{94}, 012317 (2016).

\bibitem{Mandel} L. Mandel and E. Wolf, \textit{Optical Coherence and Quantum Optics} (Cambridge University Press, 1995).

\bibitem{GH} J. Guckenheimer and P. Holmes, \textit{Nonlinear Oscillations, Dynamical Systems, and Bifurcations of Vector Fields}, (Springer,2002).

\bibitem{CohenAPI} C. Cohen-Tannoudji, J. Dupont-Roc, and G. Grynberg, \textit{Atom-Photon Iteractions: Basic Processes and Applications} (Wiley, 1992).

\bibitem{Holmes} M. H. Holmes, \textit{Introduction to Perturbation Methods} (Springer, 1995).

\bibitem{Marion} J. B. Marion and S. T. Thornton, \textit{Classical Dynamics of Particles and Systems}, 4th edition (Brooks/Cole, 1995).
 
\bibitem{Burden} R. L. Burden and J. D. Faires, \textit{Numerical Analysis}, 7th ed (Brooks/Cole, 2001).


\end{thebibliography}
\end{document}